\documentclass[12pt]{article}
\usepackage[final]{epsfig}
\usepackage{algorithm,algpseudocode}
\usepackage{amsmath}
\usepackage{makecell}
\usepackage{multicol}
\setcellgapes{8pt}
\usepackage[T1]{fontenc}
\usepackage{authblk}
\usepackage{latexsym}
\usepackage{verbatim}
\usepackage{amsthm}
\usepackage{verbatim,color}
\usepackage{graphicx}
\usepackage{amssymb}
\usepackage{amssymb, color}
\usepackage{amsmath}
\usepackage{latexsym}
\usepackage{wrapfig}
\usepackage{subfig}
\usepackage[utf8]{inputenc}
\usepackage{multicol}
\usepackage{url}
\usepackage{siunitx}
\usepackage{hyperref}
\usepackage{doi}
\usepackage{makecell}
\usepackage{enumitem}
\usepackage{xargs}                      
\usepackage[pdftex,dvipsnames]{xcolor}  
\usepackage{longtable}
\usepackage{relsize}

\makeatletter
\newenvironment{breakablealgorithm}
  {
   \begin{flushleft}
     \refstepcounter{algorithm}
     \hrule height.8pt depth0pt \kern2pt
     \renewcommand{\caption}[2][\relax]{
       {\raggedright\textbf{\fname@algorithm~\thealgorithm} ##2\par}%
       \ifx\relax##1\relax 
         \addcontentsline{loa}{algorithm}{\protect\numberline{\thealgorithm}##2}%
       \else 
         \addcontentsline{loa}{algorithm}{\protect\numberline{\thealgorithm}##1}%
       \fi
       \kern2pt\hrule\kern2pt
     }
  }{
     \kern2pt\hrule\relax
   \end{flushleft}
  }
\usepackage{array}
\usepackage{booktabs}
\usepackage{multirow}
\makeatother

\hypersetup{colorlinks=true,linkcolor=black}

\makeatletter
\def\blfootnote{\xdef\@thefnmark{}\@footnotetext}
\makeatother
\newcommand{\R}{\mathbb{R}}
\newcommand{\Z}{\mathbb{Z}}
\newcommand{\bfa}{{\bf a}}
\newcommand{\bfb}{{\bf b}}
\newcommand{\bfc}{{\bf c}}

\newcommand{\bfe}{{\bf e}}

\newcommand{\bfn}{{\bf n}}

\newcommand{\bfp}{{\bf p}}

\newcommand{\bft}{{\bf t}}
\newcommand{\bfk}{{\bf k}}

\newcommand{\bfx}{{\bf x}}
\newcommand{\bfy}{{\bf y}}

\newcommand{\bfz}{{\bf z}}
\newcommand{\bfA}{{\bf A}}

\newcommand{\bfE}{{\bf E}}
\newcommand{\bfF}{{\bf F}}

\newcommand{\bfI}{{\bf I}}

\newcommand{\bfP}{{\bf P}}
\newcommand{\bfQ}{{\bf Q}}
\newcommand{\bfR}{{\bf R}}

\newcommand{\bfT}{{\bf T}}

\newcommand{\beq}{\begin{equation}}
\newcommand{\eeq}{\end{equation}}
\newcommand{\beqs}{\begin{eqnarray}}
\newcommand{\eeqs}{\end{eqnarray}}

\newcommand{\calE}{{\cal E}}

\newcommand{\calG}{{\cal G}}

\newcommand{\calS}{{\cal S}}
\newcommand{\calT}{{\cal T}}

\newcommand{\Rmnum}[1]{\text{\uppercase\expandafter{\romannumeral #1}}}
\newtheorem{theorem}{Theorem}[section]

\newtheorem{corollary}{Corollary}[section]

\newcommand{\tr}{\operatorname{tr}}

\newcommand{\bfig}{\begin{figure}[!h]}	
\newcommand{\efig}{\end{figure}}		

\begin{document}
\bibliographystyle{unsrt}

\begin{center}
\Huge
{\bf Design of origami structures with curved
tiles between the creases} \\

\vspace{5mm}
\normalsize

\vspace{2mm}
Huan Liu and Richard D. James

\vspace{2mm}

\vspace{2mm}
Department of Aerospace Engineering and Mechanics,\\
University of Minnesota, Minneapolis, MN 55455, USA  \\
liu01003@umn.edu,  james@umn.edu

\end{center}

\vspace{0.2in}
\normalsize

\noindent {\normalsize {\bf Abstract.}  An efficient way to introduce elastic energy
that can bias an origami structure toward desired shapes is to allow curved tiles between the creases. The bending of the tiles supplies the energy and the tiles themselves may have additional functionality. In this paper, we present the theorem and systematic design methods for quite general curved origami structures that can be folded from a flat sheet, and we present methods to accurately find the stored elastic energy. Here the tiles are allowed to undergo curved isometric mappings, and the associated creases necessarily undergo isometric mappings as curves. These assumptions are consistent with
a variety of practical methods for crease design.
The $h^3$ scaling of the energy of thin sheets ($h=$ thickness) spans a broad energy range. Different tiles in an origami design can have different values of $h$, and individual tiles can also have varying $h$. Following developments for piecewise
rigid origami \cite{feng2020designs}, we develop further the Lagrangian approach and the group orbit procedure in this context.  We notice that 
some of the simplest designs that arise from the group orbit procedure for certain helical
and conformal groups provide better
matches to the buckling patterns observed
in compressed cylinders and cones than
known patterns.  
}

{\tableofcontents}

\section{Introduction}

An aspect of piecewise linear origami design appreciated by experts and recreational 
designers alike is the  huge variety of ways
a typical simple crease pattern can be folded.  This friend of
the recreational designer, and foe of the 
goal-oriented designer, is illustrated by
the (fixed) crease pattern shown in Figure  3 of \cite{liu2021origami}, having $16 \times 16$ tiles which can be folded 65,534
distinct ways, simply by varying the
mountain-valley assignment on two adjacent edges of the unfolded pseudo-rectangular sheet.

This degeneracy can potentially be removed by adding appropriate elastic
energy of the tiles by allowing them
to bend.  Great stiffness (or softness) can be achieved in this
way \cite{feng2022interfacial,tim_white}, but
little is understood how this works. In addition, by
engineering the thickness, 
tremendous freedom of the 
design of the energy
landscape is possible for a single crease pattern, simply because of the $h^3$
($h=$ thickness) dependence of energy on the thickness 
of the tiles, and the fact that different tiles can have
different thicknesses. Even on
a single tile, the thickness can
be varied smoothly with position
while preserving the property that
the tile deforms isometrically,
giving even more opportunity to design the energy landscape.

From a continuum mechanics viewpoint, the conventional approach to origami design can be
described as
Eulerian.  One looks at the deformed configuration 
and identifies kinematic objects, like angles between
neighboring tiles and distances, and then develops relationships between these qualities \cite{lang2012origami}.  An elegant version of this approach making use of the isometric transformations and concepts from
algebraic geometry is given in \cite{RJLnewbook}.

Our approach here is quite different and can be described
as Lagrangian, though we frequently make use of concepts
from the primarily Eulerian subject of differential geometry.  The goal in this case is to
give a formula for the deformation $\bfy(\bfx),\ \bfx \in \Omega$, or $\bfy(\bfx,t)$ in the dynamic case, where $\Omega$ is a flat reference configuration$-$typically a flat sheet with
a crease pattern before folding. This
approach has been developed in \cite{conti_maggi,feng2020helical,lu2022conical}; we develop further the Lagrangian method in the context of curved tile origami. We base the 
development on formulas for the rulings, and we find ways to include planar regions, inflection points, and straight segments on the crease that can occur for general
isometric mappings (see \cite{muller2005regularity, Bartels2016}). This gives quite a general Lagrangian
framework to explore curved tile
origami.  

We also further develop a group orbit method for curved tile origami structures. The general idea was outlined in \cite{ganor2016zig} as a method for constructing certain compatible microstructures arising from phase transformations in crystals and was developed for piecewise-linear origami structures in \cite{liu2021origami,feng2020helical,liu2022origami}.
This method relies on the isometry groups, which preserve
curved isometric deformations.  Different groups can be
applied to the reference configuration and the deformed
configuration, giving a variety of interesting structures. 
We further generalize this method by replacing isometry
groups by conformal groups.  These involve orthogonal transformations, translations, and dilatations.  
Fundamentally, the conformal group orbit method 
exploits the basic scaling law of 
nonlinear elasticity, $\bfy(\bfx,t) \to \eta \bfy((1/\eta)\bfx, (1/\eta)t)$, which preserves isometries, stress, equations of motion (with no body force), etc.

While exploring some of these examples,
we noticed a striking resemblance between some of our
structures to
the buckling patterns observed in 
compressed cylindrical and conical shells
(Section \ref{buckling_pattern}).
In fact, the match is apparently much better than the Yoshimura and related patterns that are
usually compared to these buckled states.  While we do not pursue
a detailed study of buckling and
bifurcation here, these patterns do 
provide explicit ``test functions'' that would be essential for such a study, and we provide methods of calculation of their energies.  It came as a surprise to us that other examples
for conformal groups show a curious resemblance to various sea creatures studied originally by Thompson
\cite{thompson1942growth}.  We do not know 
if this is purely coincidental, but we conjecture
that the group structure plays a physiological
role.

Curved tile origami structures are  of longstanding interest from the many concepts  proposed for deployable structures in space, that nevertheless have to be folded to fit into a space vehicle. But there is also rising interest from diverse areas: robotics \cite{coulson2022versatile,zhai2020situ}, foldable household and leisure items \cite{demaine2011curved}, medical devices \cite{velvaluri2021origami}, foldable buildings \cite{choma2015morphing},
wind turbines with deformable blades \cite{liu2023options} and large 
scale structures that would be otherwise difficult to transport.

\begin{longtable}[b]{p{3.5cm}<{\raggedright}p{7.5cm}<{\raggedright}<{\raggedright}p{4.5cm}}
\multicolumn{3}{l}{\small{\textbf{Table 1}} \small{Notation adopted in the paper}}\\\specialrule{0.05em}{3pt}{3pt}
Notation &Description&Formula\\\specialrule{0.05em}{3pt}{3pt}
Domain and Basis & &\\
\specialrule{0em}{1pt}{1pt}
\specialrule{0em}{1pt}{1pt}
$\hat\bfy$      &Isometric mapping&$\hat\bfy:\R^2\to\R^3$\\
\specialrule{0em}{1pt}{1pt}
$\bfx$&Point in the reference flat domain&\\
\specialrule{0em}{1pt}{1pt}
$\bfy$&Point in the deformed domain&\\
\specialrule{0em}{1pt}{1pt}
$(\hat\bfe_1,\hat\bfe_2)$& 
Fixed orthonormal basis in $\R^2$&${\rm det}(\hat\bfe_1,\hat\bfe_2)>0$\\
\specialrule{0em}{1pt}{1pt}
\multirow{1}*{$(x_1,x_2)$}&Coordinates of $\bfx$ in $(\hat\bfe_1,\hat\bfe_2)$&\multirow{1}*{$\bfx=x_1\hat\bfe_1+x_2\hat\bfe_2$}\\
\specialrule{0em}{3pt}{3pt}
Variables in crease  & &\\
\specialrule{0em}{1pt}{1pt}
$s$& Arc length parameter of creases&$s\in\R,\;s_1<s<s_2$\\
\specialrule{0em}{1pt}{1pt}
$\bfx_0$ & Reference crease &$\bfx_0(s)$\\
\specialrule{0em}{1pt}{1pt}
$\bfx_0'$ & Tangent vector of $\bfx_0$&$\bfx_0'(s)$\\
\specialrule{0em}{1pt}{1pt}
$\bfp_0$ & Principal normal vector of $\bfx_0$ in $\R^2$ &\makecell[l]{$\bfx_0'(s)\cdot\bfp_0(s)=0$, \\${\rm det}(\bfx_0'(s),\bfp_0(s))>0$} \\
\specialrule{0em}{1pt}{1pt}
$\kappa_0$ & Curvature of $\bfx_0$ &$\bfx_0''(s)\cdot\bfp_0(s)$ \\
\specialrule{0em}{1pt}{1pt}
$\bfy_0$ & Deformed crease &$\bfy_0(s)$ \\
\specialrule{0em}{1pt}{1pt}
$\bfy_0'$ & Tangent vector of $\bfy_0$&$\bfy_0'(s)$  \\
\specialrule{0em}{1pt}{1pt}
\multirow{1}*{$\bfp$} & Principal normal vector of $\bfy_0$&  \\
\specialrule{0em}{1pt}{1pt}
\multirow{1}*{$\bfb$} & Binormal vector of $\bfy_0$&\multirow{1}*{$\bfy_0'(s)\times\bfp(s)$}  \\
\specialrule{0em}{1pt}{1pt}
$\kappa$ & Curvature of $\bfy_0$ &$\bfy_0''(s)\cdot\bfp(s)$ \\
\specialrule{0em}{1pt}{1pt}
$\tau$ & Torsion of $\bfy_0$ &$-\bfb'(s)\cdot\bfp(s)$ \\
\specialrule{0em}{3pt}{3pt}
Variables in surface &\\
\specialrule{0em}{1pt}{1pt}
\multirow{1}*{$\bfy,_1,\bfy,_2$}      & Derivatives of $\hat\bfy(\bfx)$ w.r.t. $x_1$ and $x_2$&$\bfy,_\sigma=\frac{\partial\hat\bfy(\bfx(x_1,x_2))}{\partial x_\sigma},\sigma=1,2$\\
\specialrule{0em}{1pt}{1pt}
$\bfe$&Reference rulings&\\
\specialrule{0em}{1pt}{1pt}
$\bfe^\perp$&Unit vector in $\R^2$ perpendicular to $\bfe$&\makecell[l]{ $\bfe\cdot\bfe^\perp=0$,\\${\rm det}(\bfe,\bfe^\perp)$>0}\\
\specialrule{0em}{1pt}{1pt}
$\bft$&Deformed rulings&\\
\specialrule{0em}{1pt}{1pt}
$\bft^\perp$&Unit vector perpendicular to $\bft$ lying in $\bft-\bfy_0'$ plane& $\bft\cdot\bft^\perp=0,\,|\bft^\perp\cdot\bfy_0'|>0$\\
\specialrule{0em}{1pt}{1pt}
\multirow{1}*{$\bfn$}&Normal vector of a curved tile&\multirow{1}*{$\bfn=\bfy,_1\times\bfy,_2=\bft\times\bft^\perp$}\\
\specialrule{0em}{1pt}{1pt}
\multirow{1}*{$\bfF$}      &Deformation gradient of $\hat\bfy$&\multirow{1}*{$\nabla_{\bfx}\hat\bfy$}\\
\specialrule{0em}{1pt}{1pt}
$\bfe_1,\bfe_2$&Reference rulings of the two tiles meeting at $\bfx_0$ in curved tile origami&\\
\specialrule{0em}{1pt}{1pt}
$\bfe_1^\perp,\bfe_2^\perp$&Unit vectors in $\R^2$ perpendicular to $\bfe_\sigma$&\makecell[l]{ $\bfe_\sigma\cdot\bfe_\sigma^\perp=0$,\\${\rm det}(\bfe_\sigma,\bfe_\sigma^\perp)>0,\,\sigma=1,2$}\\
\specialrule{0em}{1pt}{1pt}
$\bft_1,\bft_2$&Deformed rulings of the two tiles meeting at $\bfy_0$ in curved tile origami&\\
\specialrule{0em}{1pt}{1pt}
$\bft_1^\perp,\bft_2^\perp$&Unit vectors perpendicular to $\bft_\sigma$ lying in $\bft_\sigma-\bfy_0'$ plane& \makecell[l]{$\bft_\sigma\cdot\bft_\sigma^\perp=0$, \\$|\bft_\sigma^\perp\cdot\bfy_0'|>0,\;\sigma=1,2$}\\
\specialrule{0em}{1pt}{1pt}
\multirow{1}*{$\bfn_1,\bfn_2$}&Unit normal vectors of the two tiles in curved tile origami&\\
\specialrule{0em}{1pt}{1pt}
\multirow{2}*{$\bfF_1,\bfF_2$}      &Deformation gradients of the two tiles in curved tile origami&\\
\specialrule{0em}{1pt}{1pt}
\multirow{1}*{$\gamma$}      &The angle between $\bfn_1$ and $\bfp$ &$\gamma\in(-\frac{\pi}{2}+\varepsilon,\frac{\pi}{2}-\varepsilon),\varepsilon>0$\\ 
\specialrule{0em}{1pt}{1pt}
\multirow{1}*{$\rho_1,\rho_2$}      &$C^1$ bounded functions &$\tau=\rho_1\kappa,\gamma'=\rho_2\kappa$\\ 
\specialrule{0em}{1pt}{1pt}
$f_1,f_2$      &$C^1$ bounded functions &\makecell[l]{$\tau=f_1\kappa\cos\gamma$,\\ $\gamma'=f_2\kappa\cos\gamma$}\\ 
\specialrule{0.05em}{2pt}{0pt}
\end{longtable}

\section{Necessary conditions that two isometrically deformed surfaces meet at a crease}

\subsection{Some results from differential geometry in Lagrangian form}  \label{sect2}

In this section, we collect some results from classical differential geometry for the convenience of the reader, and we rewrite some of these results in Lagrangian form.
In fact, for our constructions in the sequel, our main tools are the two formulas for the rulings in the reference and deformed 
configurations and the
invertible deformation that relates them. It will be
seen that these formulas apply in large domains (with certain restrictions) and deliver exact isometric 
mappings. For Section \ref{sect2} our presentation is informal; we work on a neighborhood of a  point on a crease with smoothness assumed as needed. 
Precise sufficient conditions that two smooth surfaces meet
at a crease are given in Section \ref{main_theorem}.

Consider a smooth deformation $ \bfy:\Omega \to \calS$ from a
domain $\Omega\subset\R^2$ into a generally curved surface $\calS = \bfy(\Omega)  \subset\R^3$.
It will be convenient to use rectangular Cartesian components $(x_1, x_2)$ relative to a fixed right-handed orthonormal basis 
$(\hat\bfe_1, \hat\bfe_2)$ for the domain $\Omega$, 
\beq
\bfx=x_1\hat\bfe_1+x_2\hat\bfe_2,\quad x_1,x_2\in\R,
\eeq 
and a description in terms of vectors for the range.

We begin by deriving some local, necessary conditions for an isometric mapping.  The deformation gradient $\bfF=\nabla_{\bfx}\bfy$ is given by
\beq
\bfF=\frac{\partial\bfy}{\partial x_i}\otimes\hat\bfe_i=\bfy_{,1}\otimes\hat\bfe_1+\bfy_{,2}\otimes\hat\bfe_2,
\label{deform_grad}
\eeq
where $\bfy_{,1}$ and $\bfy_{,2}$ are two tangent vectors of $\calS$ at $\bfy(\bfx)$. From (\ref{deform_grad}), they can be expressed as
\beq
\bfy_{,1}=\bfF\hat\bfe_1,\quad\bfy_{,2}=\bfF\hat\bfe_2.
\label{tangents}
\eeq
By definition, if $\bfy$ is an isometric mapping, it locally preserves lengths and angles in the sense that
\beqs
\bfF\bfx_1\cdot\bfF\bfx_2&=&\bfx_1\cdot\bfx_2\label{angles}
\eeqs
for all $\bfx_1,\bfx_2\in\Omega$.
Substituting (\ref{tangents}) into (\ref{angles}), we have the necessary and sufficient conditions 
\beq
\bfy_{,1}\cdot\bfy_{,1}=\bfy_{,2}\cdot\bfy_{,2}=1,\quad \bfy_{,1}\cdot\bfy_{,2}=0,
\label{eq:deform}
\eeq
for a mapping to be isometric. Clearly,
a necessary and sufficient condition for (\ref{angles}) is
\beq
\bfF^{\rm T}\bfF=\bfI.
\label{isometric}
\eeq
Since $\bfy,_i \cdot \bfy,_j = \delta_{ij}$ by (\ref{eq:deform}), we have the necessary condition for
an isometric mapping,
\beq
\bfy_{,ik}\cdot \bfy_{,j}+\bfy_{,i}\cdot \bfy_{,jk}=0.
\label{ijk}
\eeq
Firstly, let $i=j$: we get $\bfy,_{ik} \cdot \bfy,_{i}=0$ (no sum). Then let $j=k$: we get $\bfy,_{jj}\cdot \bfy,_i=0$ (no sum). So we conclude
from (\ref{ijk}) that $\bfy,_{jk} \cdot \bfy,_i=0$ for all $i,j,k=1,2$. For definiteness, we take the unit normal of $\calS$  to be $\bfn = \bfy_{,1}\times\bfy_{,2}$. Because $\bfy,_{i}$ are tangent vectors, we obtain
\beq
\bfy,_{jk}=y_{jk}\bfn,\quad j,k=1,2,\label{sec_deriv}
\eeq
where $y_{ij}=\bfy,_{ij}\cdot\bfn = y_{ji}$ are
the coefficients of the second fundamental
form.
Taking the gradient of $\bfF$, we have
\beq
\nabla\bfF=\bfy_{,ij}\otimes\hat\bfe_i\otimes\hat\bfe_j=\bfn\otimes(y_{ij}\hat\bfe_i\otimes\hat\bfe_j)
\label{Eq:gradients}
\eeq
Since $\bfy,_{ij} \cdot \bfy,_{k}=0$, we have $(\bfy,_{22} \cdot \bfy,_{1}-\bfy,_{12} \cdot \bfy,_{2}),_{1}+(\bfy,_{11}\cdot \bfy,_{2}-\bfy,_{12}\cdot \bfy,_{1}),_{2}=0.$
Expanding and simplifying above equation, we get
\beq
\bfy,_{11} \cdot \bfy,_{22}-\bfy,_{12} \cdot \bfy,_{21}=
y_{11}\, y_{22} - y_{12}^2 = 0,
\eeq
so at least one eigenvalue of the $2 \times 2$ matrix $y_{ij}$ is $0$. Hence, $y_{ij}$ is a symmetric matrix of rank one and can be written as $y_{ij}= \Lambda m_i m_j$, where $m_1^2 + m_2^2 =1$.  The two eigenvalues $0$ and $\Lambda$ are the principal curvatures of the isometrically deformed surface. Defining the unit vector
$\bfe^{\perp} = m_1 \hat\bfe_1 + m_2 \hat\bfe_2$, (\ref{Eq:gradients}) becomes 
\beq
\nabla\bfF=\Lambda\, \bfn\otimes\bfe^{\perp}\otimes\bfe^{\perp}.
\label{Eq:gradients2}
\eeq
Continuing with our local analysis, we assume by possibly shrinking $\Omega$ that $\Lambda \ne 0$ on $\Omega$\footnote{Some of our constructions below have $\Lambda=0$ in parts of the domain, but the fact that the associated mappings are exact isometric mappings will be proved from the sufficient conditions given in Theorem \ref{theorem2} below.}.  Then, $m_1, m_2$ can be chosen to be continuously differentiable on $\Omega$, and we obtain a continuously
differentiable, right-handed, orthonormal basis field $(\bfe,\bfe^\perp)$ on $\Omega$, where $\bfe = m_2 \hat\bfe_1 - m_1 \hat\bfe_2$. 

We now prove that ${\rm div}\  \bfe^\perp=0$. We begin by eliminating $\Lambda$ from (\ref{sec_deriv}) in the form
$\bfy_{ij} = \Lambda m_i m_j \bfn$.  Then we note that ${\rm div}\bfF=\bfy,_{11} + \bfy,_{22} = \Lambda \bfn$ from $m_1^2 + m_2^2 = 1$. From (\ref{Eq:gradients2}), we have $\nabla\bfF\ \bfe^\perp={\rm div}\bfF\otimes\bfe^\perp$. Taking divergence of the above equation and canceling terms, we get
\beq
\nabla\bfF\cdot\nabla\bfe^\perp=({\rm div }\ \bfe^\perp){\rm div}\bfF.
\eeq
Differentiating $\bfe^\perp \cdot \bfe^\perp = 1$, we get
$\bfe^\perp\cdot\nabla \bfe^\perp = 0$, so $\nabla\bfF\cdot\nabla\bfe^\perp=0$. Since div$\bfF = \Lambda \bfn \ne 0$, we conclude that div$\, \bfe^{\perp} = 0$ on $\Omega$.  Thus, by direct calculation, we have \beq
\nabla \bfe\, \bfe = -({\rm div}\, \bfe^{\perp}) \bfe^{\perp} = 0. \label{grade}
\eeq

\subsection{Formulas for the rulings}

We examine the integral curves of the vector field $\bfe(\bfx)$.  We choose $\bfx_0 \in \Omega$ and solve the ordinary differential equation
\beq
\bfx'(v) = \bfe(\bfx(v)), \quad \bfx(0) = \bfx_0.
\label{an_ode}
\eeq
Under our hypotheses, there is a unique local solution. Differentiating with respect to $v$, we
have that $\bfx''(v) = \nabla \bfe\, \bfe = 0$ by (\ref{grade}),
so the solution of (\ref{an_ode}) are straight
lines, $\bfx(v) = v \bfe(\bfx_0) + \bfx_0 $.  

The formulas for the rulings in the
reference domain come from the simple observation that we can allow $\bfx_0$ to lie on the smooth curve $\bfx_0(s)$ as long as it is transversal to the vector field $\bfe(\bfx)$.  Choose a smooth curve $\bfx_0(s), \ s_1 < s < s_2$, in $\Omega$ that is transversal to the vector field $\bfe(\bfx)$.  Without loss of generality, we can also assume an arc-length parameterization of this curve, so we have,
say, 
\beq
\bfx_0'(s)\cdot \bfe^{\perp}(\bfx_0(s)) > 0,\quad 
|\bfx_0'(s)| = 1, \quad s_1< s < s_2.\label{trans_x0}
\eeq
Then, for each $s\in(s_1,s_2)$, we can solve the ordinary differential equation (\ref{an_ode}) with initial condition
$\bfx(s,0) = \bfx_0(s)$ yielding the solution, 
\beq
\bfx(s,v) =\bfx_0(s)+ v \bfe(s),
\label{ref_rule}
\eeq 
where $\bfe(s) = \bfe(\bfx_0(s))$.  
The formula (\ref{ref_rule}) describes the rulings in the reference domain. Under our hypotheses, the rulings do not intersect, by uniqueness of the initial-value problem (\ref{an_ode}).  
They may not cover all of $\Omega$, but also may extend well beyond $\Omega$ in the directions $\bfe(s)$ if the deformation is defined on a larger domain.  In any case to proceed with the necessary conditions, we assume that $\Omega$ is further reduced so that the rulings are defined on all of $\Omega$.

By (\ref{Eq:gradients2}), the deformation gradient is constant along a ruling in the reference domain, i.e.,
\beq
\frac{d}{dv} \bfF(\bfx(s,v)) = \nabla \bfF(\bfx(s,v))\frac{\partial\bfx(s,v)}{\partial v} =\nabla\bfF(\bfx(s,v))\,\bfe(s)=0.
\eeq
So, with $s$ fixed, $\bfF(\bfx(s,v)) = \bfF(\bfx_0(s))$ is constant along the ruling $\bfx(s,v)$.  We consider
$\bfy(s,v) = \hat{\bfy}(\bfx(s,v))$ parameterized by the reference ruling variables $v,s$. Therefore,
we have by the chain rule
\beq
\frac{\partial}{\partial v} \bfy(s,v) = 
\bfF(\bfx(v, s))\bfe(s) = 
\bfF(\bfx_0(s))\bfe(s) =: \bft(s),
\label{deft}
\eeq
and $\bft(s)$ is a unit vector by 
(\ref{eq:deform}).
Integrating this relation, we have 
\beq
\bfy(s,v) =\bfy_0(s)+  v \bft(s).
\label{def_rule}
\eeq
Since $\bfy_0(s)$ is the isometric image
of the arclength parameterized curve $\bfx_0(s)$, then it is also parameterized by arclength with $|\bfy_0'(s)|=1$.

Additional necessary conditions
on the rulings follow from the fact that
$\bfF(s) := \bfF(\bfx_0(s))$ is constant
on a ruling.  Differentiating $\hat{\bfy}(\bfx(s,v)) = \bfy(s,v)$ with respect to
$s$, we have the condition
\beq
v \big(\bfF(s) \bfe'(s) - \bft'(s)\big) = \bfy_0'(s) - \bfF(s) \bfx_0'(s),
\eeq
which, combined with (\ref{deft}), implies
that
\beq
\bfF(s)\bfe(s) = \bft(s), \quad 
\bfF(s)\bfx_0'(s) = \bfy_0'(s), \quad 
\bfF(s)\bfe'(s) = \bft'(s)
\label{3conds}
\eeq
on $\Omega$.  Define $\bft^\perp(s)$ by $\bft^\perp(s)=\bfn(s)\times\bft(s)$. Then $\bfx_0'(s)$ and $\bfy_0'(s)$ can be expressed as
\beq
\begin{array}{ccc}
    \bfx_0'(s)&= & (\bfx_0'\cdot\bfe)\bfe(s)+ (\bfx_0'\cdot\bfe^\perp)\bfe^\perp(s),\\
    \bfy_0'(s)&= &(\bfy_0'\cdot\bft) \bft(s)+ (\bfy_0'\cdot\bft^\perp)\bft^\perp(s). 
\end{array}\label{x0'-y0'}
\eeq
We use the first and second of
(\ref{3conds}), (\ref{x0'-y0'}) and the fact that $\bfF^{\rm T}\bfF=\bfI$ to get
\beqs
&\bfx_0'\cdot\bfe=\bfy_0'\cdot\bft,\quad\bfx_0'\cdot\bfe^\perp=\bfy_0'\cdot\bft^\perp,&\label{x0_e_y0_t}\\
&\bfF(s) = \bft(s) \otimes \bfe(s) +
 \bft^\perp(s)  \otimes \bfe^{\perp}(s).&   \label{preF(s)}
\eeqs
Finally, the third condition of (\ref{3conds}) applied to (\ref{preF(s)}) gives
\beq
\bft'(s) = (\bfe^{\perp}(s) \cdot \bfe'(s))\ \bft^\perp(s),
\label{another_ode}
\eeq
which implies 
\beq
\bft'(s) \cdot (\bft(s) \times \bft^{\perp}(s)) = 0.\label{necessary_t}
\eeq

\subsection{Formulas for the creases}
\label{FS}
In our main Theorem \ref{theorem2} below we use the  Frenet-Serret formulas to define the crease $\bfy_0(s),s_1<s<s_2$. The tangent $\bfy_0'(s)$, principal normal $\bfp(s)$ and binormal $\bfb(s)$ of a smooth 
arclength-parameterized curve $\bfy_0(s)$ with (signed) curvature $\kappa(s)$ and torsion $\tau(s)$ satisfy
the linear system of ordinary differential equations
\beq
\begin{array}{cclll}
\bfy_0''(s) &=& \;\;\,\kappa(s) \bfp(s), &&\bfy_0'(s_1) = \bar\bfy'_{0}, \\
\bfp'(s) &=& -\kappa(s) \bfy_0'(s) + \tau(s) \bfb(s),&&\bfp(s_1) =\bar\bfp,\\
\bfb'(s) &=& -\tau(s) \bfp(s),&&\bfb(s_1) =\bar\bfb.
\end{array}
\label{fs}
\eeq
(The precise smoothness and other conditions on $\kappa(s)$ and $\tau(s)$ are given in Theorem \ref{theorem2}.)
If the initial conditions $(\bar\bfy_{0}', \bar\bfp, \bar\bfb)$ are right-handed orthonormal, then the functions $\bfy_0', \bfp, \bfb$ remain orthonormal for $s_1< s< s_2$ due to the the fact that their
pairwise dot products satisfy a system of linear ordinary differential equations (ODEs) in standard form with initial
conditions
\beq
(|\bfy_0'|^2, |\bfp|^2, |\bfb|^2,\ \bfy_0' \cdot \bfp,\ \bfy_0' \cdot \bfb,\ \bfp \cdot \bfb)|_{s = s_1} = (1,1,1,0,0,0).  \label{111000}
\eeq
By direct observation, these equations continue to have the solution $(1,1,1,0,0,0)$ for $s> s_1$.  The right-handedness follows from the right-handedness of the initial data and the continuity of this solution.

We emphasize that, unlike typical books on differential geometry, we do not assume that the curvature (or the torsion) is non-negative, but we allow it to have both signs in Section \ref{main_theorem}.  Then the Frenet-Serret formulas make sense as ordinary differential
 equations and give curves that may have complex  arrays of inflection points or straight regions. 
  This is essential: many of our designs below have creases with inflection points.  Similarly, the normal and
 binormal are also continuously differentiable.
 The curve with (signed) curvature $\kappa(s)$ and torsion $\tau(s)$ can
be obtained by integrating the tangent,
\beq
 \bfy_0(s) = \bar\bfy_0 + \int_{s_1}^s \bfy_0'(r)\, dr.
 \eeq

A smooth reference crease $\bfx_0(s),s_1<s<s_2$ in $\R^2$ can be defined by using two-dimensional Frenet-Serret formulas. That is, the tangent $\bfx_0'(s)$ and the principal normal $\bfp_0(s)$ with (signed) curvature $\kappa_0(s)$ satisfy
\beq
\begin{array}{cclll}
\bfx_0''(s) &=& \;\;\,\kappa_0(s) \bfp_0(s), &&\bfx_0'(s_1) = \bar\bfx'_{0}, \\
\bfp_0'(s) &=& -\kappa_0(s) \bfx_0'(s),&&\bfp_0(s_1) =\bar\bfp_0,
\end{array}
\label{fs0}
\eeq
with $\bar\bfx_0'\cdot\bar\bfp_0=0$. Then, $\bfx_0'(s), \bfp_0(s)$ remain 
orthonormal on $(s_1, s_2)$. $\bfx_0(s)$ can be obtained  by
\beq
 \bfx_0(s) = \bar\bfx_0 + \int_{s_1}^s \bfx_0'(r)\, dr.
 \eeq
In the following contents, we consider $\bfy_0(s)$ and $\bfx_0(s)$ are curves that can be defined by (\ref{fs}) and (\ref{fs0}), respectively. 

\subsection{Formulas for isometrically deformed curved origami}

In this section, we collect some necessary conditions for curved tile origami such that the whole structure is developable,
i.e., isometric to a subset of a plane.
First of all, 
consider a curve $\bfy_0(s),s_1<s<s_2$ lying on a curved developable surface $\bfy(s,v),s_1<s<s_2,v_1<v<v_2$ with  normal  $\bfn(s)$. Let $\bfx_0(s)$ be the preimage of $\bfy_0(s)$. Since $\bfy_0'=\bfF\bfx_0'$ and $\bfy_0''=\bfF\bfx_0''+\nabla\bfF\bfx_0'\bfx_0'$, by substituting (\ref{deform_grad}) and (\ref{Eq:gradients2}) we get $\bfy_0'\cdot\bfy,_\sigma=\bfx_0'\cdot\hat\bfe_\sigma$ and  $\bfy_0''\cdot\bfy,_\sigma=\bfx_0''\cdot\hat\bfe_\sigma,\sigma=1,2$. Then
\beqs 
\kappa \bfn \cdot \bfb&=&(\bfy,_1\times\bfy,_2)\cdot(\bfy_0'\times\bfy_0'')\nonumber\\
&=&(\bfy_0'\cdot\bfy,_1)(\bfy_0''\cdot\bfy,_2)-(\bfy_0''\cdot\bfy,_1)(\bfy_0'\cdot\bfy,_2)\nonumber\\
&=&(\bfx_0'\cdot\hat\bfe_1)(\bfx_0''\cdot\hat\bfe_2)-(\bfx_0''\cdot\hat\bfe_1)(\bfx_0'\cdot\hat\bfe_2)\nonumber\\
&=&\bfx_0''\cdot\bfp_0=\kappa_0.\label{n.b}
\eeqs
Since $\bfn\cdot\bfy_0'=0$, $\bfn$ can be expressed as
\beq
\bfn=\pm\cos\gamma\bfp+\sin\gamma\bfb,
\eeq
where $\gamma\in(-\pi/2 ,\pi/2)$ satisfies 
\beq
\kappa_0=\kappa\sin\gamma.\label{k0-k}
\eeq

Now consider two generally curved surfaces with distinct normal vectors $\bfn_1$ and $\bfn_2$ join at a curve $\bfy_0\in C^2$. Assume the whole structure is isometric to a plane without overlapping, and let $\bfx_0$ be the preimage of $\bfy_0$. Then (\ref{n.b}) shows that $\kappa
\bfn_1 \cdot \bfb=\kappa
\bfn_2 \cdot \bfb=\kappa_0$. Without loss and generality, $\bfn_1$ and $\bfn_2$ can be given by
\beq
\bfn_{1}=\cos\gamma\, \bfp+\sin\gamma\bfb ,\quad\bfn_{2}=-\cos\gamma\bfp+\sin\gamma\, \bfb.\label{kappa0-1}
\eeq
(At  straight crease segments, $\bfp$ and $\bfb$ can be specified in such a way that (\ref{kappa0-1}) remains satisfied.) Thus, 
\beq
\bfn_2=(-\bfI+2\bfb\otimes\bfb)\bfn_1. \label{developability}
\eeq
According to (\ref{k0-k}) and (\ref{kappa0-1}), if $\bfy_0$ is given (i.e., $\kappa,\bfp$ and $\bfb$ are given), for each $\bfx_0$ (i.e., $\kappa_0$), there are only two developable surfaces with normal vetors $\bfn_1,\bfn_2$ that can go through $\bfy_0$. If we assign a different reference crease $\bfx_0$, we can get different pairs of developable surfaces.

To avoid a trivial folding (cf., Figure \ref{fig:four_branches}(b) below), we assume $\bfn_1$ and $\bfn_2$ are distinct, i.e., $|\bfn_1\cdot\bfn_2|\le 1-2\varepsilon^2<1$ for some $\varepsilon>0$. Since $\bfn_1\cdot\bfn_2=\bfn_1\cdot(-\bfI+2\bfb\otimes\bfb)\bfn_1=2(\bfb\cdot\bfn_1)^2-1\le 1-2\varepsilon^2$
and $\bfb\cdot\bfn_1=\bfb\cdot\bfn_2$, we have $|\bfb\cdot\bfn_\sigma|^2\le1-\varepsilon^2<1, \sigma=1,2$. Via (\ref{kappa0-1}) this is equivalent to \beq
|\bfp\cdot\bfn_\sigma|\ge\varepsilon>0,\quad\sigma=1,2.
\label{pn}
\eeq

Let $\bft_1$ and $\bft_2$ denote the rulings of the two surfaces.  We  assume transversality, i.e., that the ruling $\bft_\sigma(s)$ and tangent to the crease $\bfy_0'(s)$ do not become parallel: 
\beq
|\bft_\sigma^\perp\cdot\bfy_0'|\geq c>0,\quad\sigma=1,2,\label{tranversality}
\eeq
for some $0< c \ll 1$. This appears to be quite natural in the case of origami design.
Since $\bft_\sigma\cdot\bfn_\sigma=0$, $\bft_1,\bft_2$ can be expressed as
\beq
\begin{array}{ccl}
\bft_1 &=& \cos \alpha_1 \, \bfy_0' - \sin \alpha_1 \,  \bfn_1 \times \bfy_0' , \\
\bft_1^{\perp} &=& \sin \alpha_1 \, \bfy_0' + \cos \alpha_1 \, \bfn_1 \times \bfy_0',
\end{array}\quad \begin{array}{ccl}
\bft_2 &=& \cos \alpha_2 \, \bfy_0' - \sin \alpha_2 \,  \bfn_2 \times \bfy_0' , \\
\bft_2^{\perp} &=& \sin \alpha_2 \, \bfy_0' + \cos \alpha_2 \, \bfn_2 \times \bfy_0',
\end{array}
\eeq
where $\alpha_1\in(-\pi + \hat{c},-\hat{c})$ and $\alpha_2\in(\hat{c},\pi - \hat{c})$, $\hat{c} = \arcsin c$.
The angles $\alpha_\sigma$ represent the angles between the rulings $\bft_\sigma$ and $\bfy_0'$. Then $\alpha_1,\alpha_2$ satisfy the necessary conditions $\bft_1' \cdot \bfn_1 = 0$ and $\bft_2' \cdot \bfn_2 = 0$ from (\ref{necessary_t}), which imply
\beq
\kappa\cos\gamma\cot \alpha_1=\tau+\gamma',\quad\kappa\cos\gamma\cot \alpha_2=-\tau+\gamma'.\label{gamma-tau}
\eeq
Necessary and sufficient conditions that there exist bounded $C^1$ functions $\cot\alpha_1$ and $\cot\alpha_2$ are that there exist bounded $C^1$ functions $\rho_1,\rho_2$ satisfying 
\beq
\tau=\rho_1\kappa,\quad\gamma'=\rho_2\kappa.
\eeq
Here, we have also used (\ref{pn}).

\section{Theorem on curved origami design}
\label{main_theorem}
In this section, we present the main theorem of the paper.  We consider the classic problem of origami design of two smooth surfaces
meeting at a deformed crease, and we wish to design the crease pattern in the
reference domain, i.e., a single flat sheet.   These two smooth surfaces must be 
obtainable from isometric deformations of this reference domain. 
One would then like to predict the reference 
crease that deforms isometrically to the given deformed crease and the reference rulings that deform 
to the given deformed rulings.  The latter can guide the folding process.  

Under mild conditions of smoothness, this is
easily accomplished for just one surface, and there are many such surfaces.  (This is obvious 
because one can take any isometric mapping of a flat sheet and simply draw a curve on the deformed surface.  Taking
the inverse image of this curve gives a reference crease.)   However, the addition of a second surface meeting 
the same crease becomes quite restrictive: under mild restrictions, the first surface and the deformed crease 
determine the second surface. 

Under mild restrictions, Theorem \ref{theorem2} below treats two isometrically deformed surfaces meeting at a deformed crease, both obtained by isometric 
deformations of a flat sheet and both sharing the same reference crease.

\begin{theorem} \label{theorem2} Let 
curvature and torsion, $\kappa, \tau \in C^1(s_1, s_2)$ of the deformed crease  be given satisfying $\tau=\rho_1\kappa$, where $\rho_1$ is a bounded $C^1$ function.  Let the
deformed crease $\bfy_0 \in C^2(s_1, s_2)$ together with its principal normal and binormal $\bfp,\bfb \in C^2(s_1, s_2)$  be the unique solutions of the Frenet-Serret equations
\beq 
\begin{array}{ccl} \bfy_0''&=&\;\;\,\kappa\bfp, \quad \quad \quad \quad \bfy_0'(s_1) = \bar{\bfy}_0',\\
\bfp'&=&-\kappa\bfy_0'+\tau\bfb, \quad \ \ \bfp(s_1) = \bar{\bfp},\\
\bfb'&=&-\tau\bfp, \quad  \quad \quad \quad \  \, \bfb(s_1) = \bar{\bfb},\\ 
\end{array}
\label{fs_thm3.1}
\eeq
with given right-handed orthonormal  initial values $\bar{\bfy}_0', \bar{\bfp}, \bar{\bfb}$. (Alternatively, give the
deformed crease $\bfy_0$ with the indicated smoothness having curvature and torsion satisfying $\tau=\rho_1\kappa$ and calculate $\bfp, \bfb$ consistent with the Frenet-Serret equations.)
We restrict the domain of $\bfy_0$, if necessary,
so that it does not intersect itself on $[s_1, s_2]$.

To define Surface 1 and  Surface 2, let $\gamma(s)\in[-\frac{\pi}{2}+\varepsilon,\frac{\pi}{2}-\varepsilon]$, $\varepsilon>0$, be a $C^1(s_1, s_2)$ function satisfying $\gamma'=\rho_2\kappa$, where $\rho_2$ is a bounded $C^1(s_1, s_2)$ function. Let
\beq
\begin{array}{ccl}
\bfn_1&=& \cos\gamma\bfp+\sin\gamma\bfb,\\
\bft_1 &=& \cos \alpha_1 \, \bfy_0' - \sin \alpha_1 \,  \bfn_1 \times \bfy_0' ,\\
\bft_1^{\perp} &=& \sin \alpha_1 \, \bfy_0' + \cos \alpha_1 \, \bfn_1 \times \bfy_0',
\end{array}\quad
\begin{array}{ccl}
\bfn_2 &=& -\cos\gamma\bfp+\sin\gamma\bfb,\\
\bft_2 &=& \cos \alpha_2 \, \bfy_0' - \sin \alpha_2 \,  \bfn_2 \times \bfy_0' ,\\
\bft_2^{\perp} &=& \sin \alpha_2 \, \bfy_0' + \cos \alpha_2 \, \bfn_2 \times \bfy_0',
\end{array}\label{surf1_surf2}
\eeq
where $\alpha_1\in(-\pi,0),\alpha_2\in(0,\pi)$ are defined by
\beq
\alpha_1=\cot^{-1}\left(\frac{\rho_1+\rho_2}{\cos\gamma}\right),\quad\alpha_2=\cot^{-1}\left(\frac{\rho_2-\rho_1}{\cos\gamma}\right).  \label{alphas}
\eeq
Define the reference crease $\bfx_0$  by the following ODEs:
\beq
\left\{\begin{array}{ccll}
\bfx_0''&=&\;\;\,\kappa_0\bfp_0,&\bfx_0'(s_1)=
\bar\bfx_0',\\
    \bfp_0'&=&-\kappa_0\bfx_0',&\bfp_0(s_1)=\bar\bfp_0,
\end{array}\right. \label{refcr_Thm3.1}
\eeq 
where 
\beq
\kappa_0=\kappa\bfn_\sigma\cdot\bfb=\kappa\sin\gamma,\quad \sigma=1,2.\label{reference_curvatures}
\eeq The reference rulings of Surface 1 and Surface 2 are then defined by
\beqs
\bfe_\sigma&=&\cos\alpha_\sigma\, \bfx_0'-\sin\alpha_\sigma\,\bfp_0,\label{rsurf1_rsurf2_1}\\
\bfe_\sigma^\perp&=&\sin\alpha_\sigma\, \bfx_0'+\cos\alpha_\sigma\, \bfp_0,\quad\sigma=1,2.\label{rsurf1_rsurf2_2}
\eeqs
The parameterizations of Surfaces 1 and 2 in terms of creases and rulings are
\beqs
\bfx_{\sigma}(s,v)&=&\bfx_0(s)+v\bfe_\sigma(s),\\
\bfy_{\sigma}(s,v)&=&\bfy_0(s)+v\bft_\sigma(s), \ \ \sigma = 1, 2
\eeqs
respectively, with $(s,v) \in \Omega_{\sigma}$, where $\Omega_1=\{(s,v):s_1 < s < s_2,-v_1^-(s) < v < v_1^+(s)\}$, $\Omega_2=\{(s,v):s_1 < s < s_2,
-v_2^-(s) < v < v_2^+(s)\}$, and, by the restrictions on $\alpha_1$ and $\alpha_2$, $v_1^-(s),\dots,v_2^+(s)>0$ on $(s_1, s_2)$ are assigned such that there are no intersections between rulings. 

Let $\hat{\bfy}_{\sigma}(\bfx)$, $\sigma = 1, 2, $ be the induced mappings between
rulings:
\beq
\hat{\bfy}_{\sigma}(\bfx_{\sigma}(s,v)) = \bfy_{\sigma}(s,v), \quad (s, v) \in \Omega_{\sigma}. 
\eeq
 Then the two mappings $\hat{\bfy}_1$ and $\hat{\bfy}_2$ are each isometric.  
\end{theorem}
\begin{figure}[ht]
\centering
\includegraphics[scale=1]{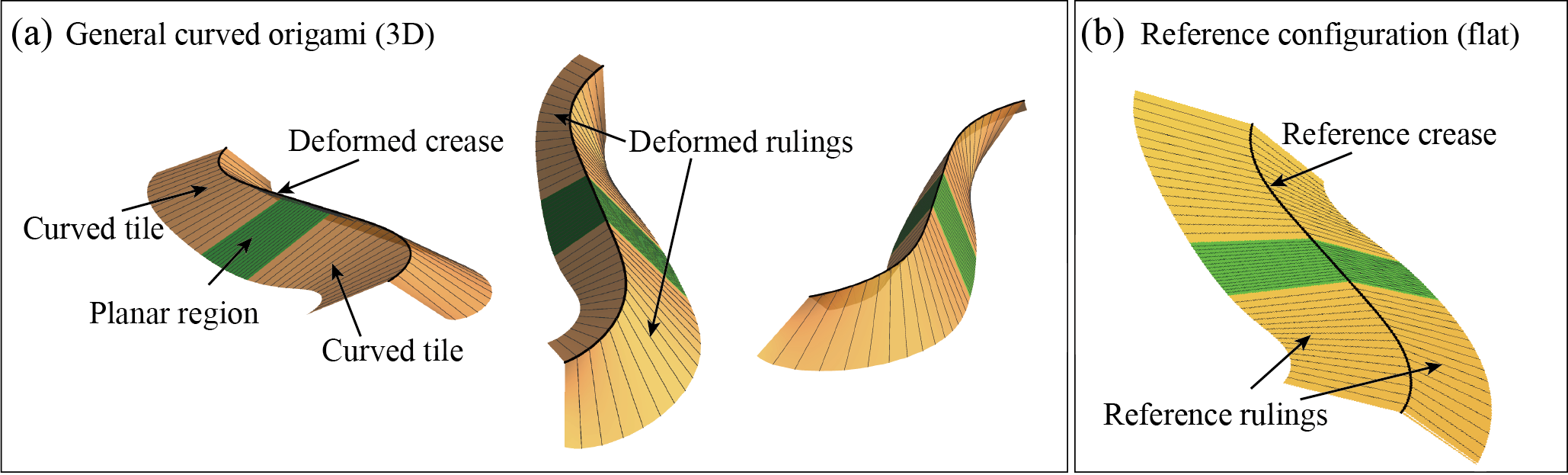}
\caption{General curved origami given by Theorem \ref{theorem2} having planar regions (green). (a) Deformed configuration from three different viewpoints. (b) The  reference configuration of (a). The surfaces highlighted in green in (a-b) are planar regions meeting at a straight crease segment.   Note that the sheet in (b) is flat: the apparent bending at the crease is an optical illusion.} 
\label{fig:continuous_normal}
\end{figure}
\begin{proof}
The formula (\ref{surf1_surf2}) implies that $\{\bft_\sigma,\bft_\sigma^\perp,\bfn_\sigma\},\sigma=1,2$ form two right-handed orthonormal bases. The formulas (\ref{alphas}) are well defined since $\rho_1,\rho_2,\cos\gamma$ are bounded functions and the cotangent is invertible on $(-\pi,0)$ and on $(0,\pi)$.
 
The invertibility of $\bfx_{\sigma} (s,v)$ on
$\Omega_{\sigma}, \sigma = 1, 2$ follows from the
condition that the rulings do not intersect.  By the smoothness and invertibility of $\bfx_{\sigma} (s,v)$, we have that $\hat{\bfy}_{\sigma}(\bfx_\sigma) \in C^1(\Omega_{\sigma})$.  Let the deformation gradient of $\hat\bfy_1$ be $\bfF_1=\nabla_{\bfx_1}\hat\bfy_1$. Then $\bfF_1$ satisfies
\beq
\bfy_0'+v\bft_1'=\bfF_1(\bfx_0'+v\bfe_1'),\quad\bft_1=\bfF_1\bfe_1.\label{proof_F1}
\eeq
According to (\ref{surf1_surf2}-\ref{alphas}), $\bft_1',\bfe_1'$ are found by $\bft_1'=(\kappa_0-\alpha_1')\bft_1^\perp$ and  $\bfe_1'=(\kappa_0-\alpha_1')\bfe_1^\perp$. Substituting to the first of (\ref{proof_F1}), and since (\ref{proof_F1}) is true for all $v\in(-v_1^-,v_1^+)$, one can get
\beq
\bfy_0'-\bfF_1\bfx_0'=0,\quad(\kappa_0-\alpha_1')(\bft_1^\perp-\bfF_1\bfe_1^\perp)=0.\label{proof_F1-2}
\eeq
From (\ref{surf1_surf2}) and (\ref{rsurf1_rsurf2_1}-\ref{rsurf1_rsurf2_2}),  $\bfy_0',\bfx_0'$ can be expressed as
\beq
\bfy_0'=\cos\alpha_1\bft_1+\sin\alpha_1\bft_1^\perp,\quad\bfx_0'=\cos\alpha_1\bfe_1+\sin\alpha_1\bfe_1^\perp.\label{y0'-x0'}
\eeq
Substitute (\ref{y0'-x0'}) into the first of (\ref{proof_F1-2}) and use the restriction on the domain of $\alpha_1$ to get
\beq
\bft_1^\perp=\bfF_1\bfe_1^\perp,
\eeq
which, together with the second of (\ref{proof_F1}) ($\bft_1=\bfF_1\bfe_1$), gives
\beq
\bfF_1=\bft_1\otimes\bfe_1+\bft_1^\perp\otimes\bfe_1^\perp.\label{deformation_gradient-1}
\eeq
Since $\bfF_1$ satisfies $\bfF_1^{\rm T}\bfF_1=\bfI$ (cf., (\ref{isometric})), the mapping $\hat\bfy_1:\bfx_1(\Omega_1)\to\bfy_1(\Omega_1)$ is isometric. Similarly, the deformation gradient for $\hat\bfy_2$ can be found by
\beq
\bfF_2=\bft_2\otimes\bfe_2+\bft_2^\perp\otimes\bfe_2^\perp,\label{deformation_gradient-2}
\eeq
which implies the mapping $\hat\bfy_2:\bfx_2(\Omega_2)\to\bfy_2(\Omega_2)$ is also isometric.
\end{proof}
In brief, if $\bfy_0$ and the normal vector $\bfn_1$ of one surface are well defined under the assumptions in the theorem, this surface will be completely determined. The other surface is then determined by using its normal vector $\bfn_2=(-\bfI+2\bfb\otimes\bfb)\bfn_1$ (see an example in Figure \ref{fig:four_branches}(a-b)).  Specifying different $\bfn_1$, one will get different pairs of surfaces, and the preimage $\bfx_0$ will change correspondingly.
To get nontrivial curved origami structures, we assign the two mappings to each side of the crease, which gives two distinct deformed configurations, see Figure \ref{fig:four_branches}(c).

\begin{figure}[H]
\centering
\includegraphics[scale=1]{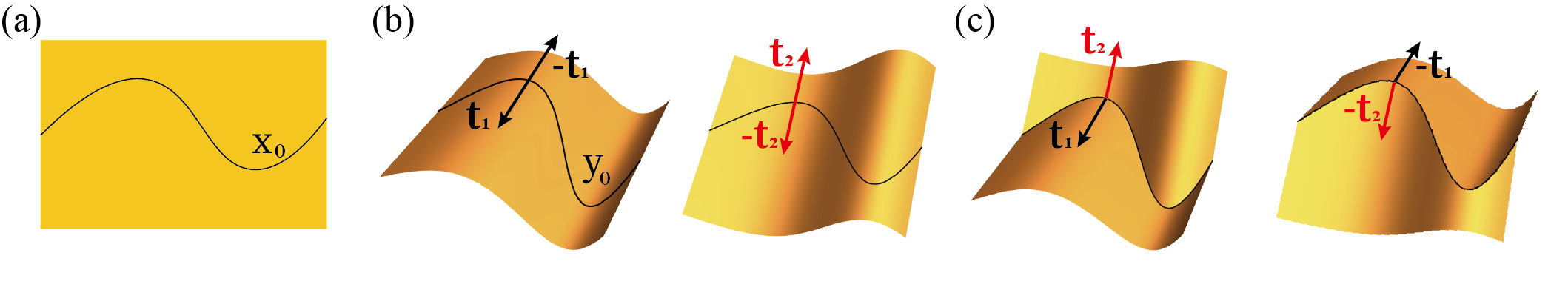}
\caption{Examples of curved tile origami given by Theorem \ref{theorem2}. (a) Reference domain. (b) Surface 1 and Surface 2 isometrically deformed from (a) going through $\bfy_0$. The ruling directions on both sides of $\bfy_0$ are: $(\bft_1,-\bft_1)$ and $(-\bft_2,\bft_2)$. (c) Two curved origami structures by assigning these two mappings to different sides of the crease. The ruling directions on opposite sides of $\bfy_0$ are: $(\bft_1,\bft_2)$ and $(-\bft_2,-\bft_1)$.}
\label{fig:four_branches}
\end{figure}

\begin{figure}[ht]
\centering
\includegraphics[scale=1]{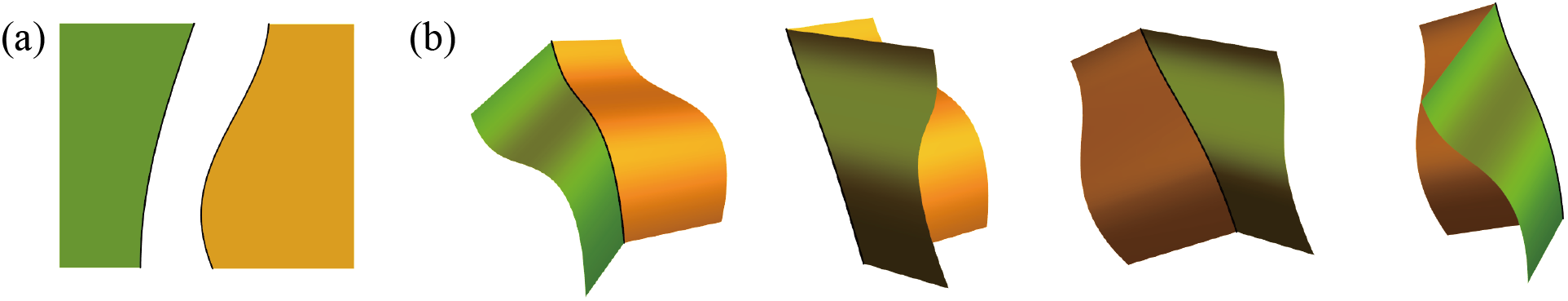}
\caption{An example of generalized curved origami. (a) Incompatible reference domain. (b) The four branches of deformed configurations.}
\label{fig:general_case}
\end{figure}

In Theorem \ref{theorem2}, one can assign $\kappa, \tau$
and the initial conditions in (\ref{fs_thm3.1}) to get a unique deformed
crease $\bfy_0$. Then different choices of $\gamma$ will define different $\kappa_0$ (see \ref{reference_curvatures}), which give different reference creases $\bfx_0$. Here we make two different choices of $\gamma(s)$, say
$\gamma_1(s), \gamma_2(s)$, 
to get altogether four surfaces passing
through the crease $\bfy_0$.  Among these four,
select one surface corresponding to $\gamma_1$ and another corresponding to
$\gamma_2$.  In this way, we construct two
different reference creases that deform
to the same deformed crease and, correspondingly, two isometric mappings.
This method applies to the case where the two reference regions are not compatible as shown in Figure \ref{fig:general_case}(a). This case is exploited in famous architectural designs
such as the Walt Disney Concert Hall in 
Los Angeles, California. Assume the two reference creases and the deformed crease are well prescribed. Then one can join two incompatible flat sheets with different creases together at the same deformed crease by solving each pair of reference crease and deformed crease individually. See Figure \ref{fig:general_case} for the four
possibilities of this type.

In summary, our methods contained in Theorem \ref{theorem2} 
treat quite generally cases analogous to classical origami, i.e., piecewise isometric folding a creased sheet (Figure \ref{fig:continuous_normal} and Figure \ref{fig:four_branches}), or cases seen in Frank Gehry designed buildings (Figure \ref{fig:general_case}) in which sheets
bounded by two different reference creases are deformed isometrically to join at the same deformed crease.

The following series of corollaries follow immediately
from Theorem \ref{theorem2}.

\begin{corollary}
\label{normal_grad}
Assume the hypotheses of Theorem \ref{theorem2}.
From (\ref{surf1_surf2}), the normal vectors $\bfn_1$ and $\bfn_2$ satisfy
\beq
\bfn_1=-\bfP\bfn_2.
\eeq
 where $\bfP=\bfI-2\bfb\otimes\bfb$.   Substituting $\bft_\sigma,\bft_\sigma^\perp$ and $\bfe_\sigma,\bfe_\sigma^\perp$ into $\bfF_\sigma,\sigma=1,2$, $\bfF_1$ and $\bfF_2$ satisfy
    \beq
    \bfF_1=\bfP\bfF_2,
    \eeq
i.e., the two deformation gradients $\bfF_1$ and $\bfF_2$ given by Theorem \ref{theorem2} are related by a reflection through a plane having normal given by the binormal of the crease.
\end{corollary}

\begin{corollary}
    Let $\bfy_0$ and $\gamma$ be prescribed consistent with Theorem \ref{theorem2}.  Explicit formulas of quantities given by Theorem \ref{theorem2} for constructing curved origami are collected in Table \ref{formulas0}. In the table, $f_1(s)$ and $f_2(s)$ are
\beq
f_1=\frac{\rho_1}{\cos\gamma},\quad f_2=\frac{\rho_2}{\cos\gamma},
\eeq
that is, $f_1$ and $f_2$ are bounded $C^1$ functions satisfying
\beq
\tau=f_1\ \kappa\cos\gamma,\quad\gamma'=f_2\ \kappa\cos\gamma.
\eeq
\vspace{-3mm}
\begin{table}[h]
\makegapedcells
\caption{Formulas of constructing curved origami}
\vspace{-5mm}
\label{formulas0}
\begin{center}
\begin{tabular}{lcc} 
\Xhline{2\arrayrulewidth}
\textbf{Variable} & \multicolumn{2}{c}{\textbf{Crease}}\\ 
\Xhline{2\arrayrulewidth}
$\bfx_0$& \multicolumn{2}{c}{$\int_{s_1}^s\ \left(\cos(\int_{s_1}^{\hat s}\kappa\sin\gamma\ dr) \hat\bfe_1+\sin(\int_{s_1}^{\hat s}\kappa\sin\gamma\ dr) \hat\bfe_2\right)\ d\hat s
$}    \\
\Xhline{0.5\arrayrulewidth}
$\bfx_0'$& \multicolumn{2}{c}{$\cos(\int_{s_1}^s\kappa\sin\gamma\ dr) \hat\bfe_1+\sin(\int_{s_1}^s\kappa\sin\gamma\ dr) \hat\bfe_2
$}    \\
\Xhline{0.5\arrayrulewidth}
$\bfp_0$& \multicolumn{2}{c}{$-\sin(\int_{s_1}^s\kappa\sin\gamma\ dr) \hat\bfe_1+\cos(\int_{s_1}^s\kappa\sin\gamma\ dr) \hat\bfe_2
$}    \\
\Xhline{2\arrayrulewidth}
\textbf{} & \textbf{Surface 1}& \textbf{Surface 2}\\ 
\Xhline{2\arrayrulewidth}
$\bfn$ & $ \cos\gamma\bfp+ \sin\gamma\bfb
$& $ -\cos\gamma\bfp+ \sin\gamma\bfb
$\\ 
\Xhline{0.5\arrayrulewidth}
$\bfn\times\bfy_0'$ & $ \sin\gamma\bfp-\cos\gamma\bfb$& $\sin\gamma\bfp+\cos\gamma\bfb
$\\ 
\Xhline{0.5\arrayrulewidth}
$\bft^\perp$   &$\displaystyle{\frac{-\bfy_0'-(f_1+f_2)\bfn_1\times\bfy_0'}{\sqrt{(f_1+f_2)^2+1}}}$&$\displaystyle{\frac{\bfy_0'-(f_1-f_2)\bfn_2\times\bfy_0'}{\sqrt{(f_1-f_2)^2+1}}}$\\
\Xhline{0.5\arrayrulewidth}
$\bft$    &$\displaystyle{\frac{-(f_1+f_2)\bfy_0'+\bfn_1\times\bfy_0'}{\sqrt{(f_1+f_2)^2+1}}}$&$\displaystyle{\frac{-(f_1-f_2)\bfy_0'-\bfn_2\times\bfy_0'}{\sqrt{(f_1-f_2)^2+1}}}$\\ 
\Xhline{0.5\arrayrulewidth}
$\bfe^\perp$      &$\displaystyle{\frac{-\bfx_0'-(f_1+f_2)\bfp_0}{\sqrt{(f_1+f_2)^2+1}}}$&$\displaystyle{\frac{\bfx_0'-(f_1-f_2)\bfp_0}{\sqrt{(f_1-f_2)^2+1}}}$\\ 
\Xhline{0.5\arrayrulewidth}
$\bfe$    &$\displaystyle{\frac{-(f_1+f_2)\bfx_0'+\bfp_0}{\sqrt{(f_1+f_2)^2+1}}}$&$\displaystyle{\frac{-(f_1-f_2)\bfx_0'-\bfp_0}{\sqrt{(f_1-f_2)^2+1}}}$\\ 
\Xhline{2\arrayrulewidth}
\end{tabular}
\end{center}
\end{table}
\vspace{-5mm}
\end{corollary}
\begin{corollary}
\label{planar_crease}
    If $\bfy_0$ is a planar curve, then $\bfb$ is constant and $\tau=0$ on $(s_1,s_2)$. Thus, we can choose $f_1=0$ in Theorem \ref{theorem2}. According to Table \ref{formulas0}, the rulings satisfy
    \beq
    \bft_1=-\bfP\bft_2,\quad\bfe_1=-\bfe_2.
    \eeq
\end{corollary}
\begin{corollary}
\label{theorem3}
In addition to constructing curved origami by specifying the deformed crease and one surface or by specifying the reference crease and the deformed crease, one can also specify the two distinct normal vectors $\bfn_1,\bfn_2$. Let $\bfn_1,\bfn_2\in C^1$ be distinct normal vectors of the two surfaces satisfying $\bfn_\sigma'=\tau_\sigma\bft_\sigma^\perp$, where $\tau_\sigma, \bft_\sigma^\perp\in C^1,\sigma=1,2$. Since $\bfy_0'\cdot\bfn_1=0$ and $\bfy_0'\cdot\bfn_2=0$, the crease $\bfy_0$ is given by
    \beq
\bfy_0(s)=\bar\bfy_0+\int_{s_1}^s\frac{\bfn_{1}\times\bfn_{2}}{|\bfn_{1}\times\bfn_{2}|}dr.\label{theorem3_y0}
\eeq
where $\hat\bfy_0\in\R^3$. The transversality condition and the developability condition imply $\bfn_1$ and $\bfn_2$ satisfy
\beq
|(\bfn_1\times\bfn_2)\cdot\bft_\sigma^\perp|>0,\quad(\bfn_{+1}\times\bfn_{-1})\cdot(\bfn_{+1}'+\bfn_{-1}')=0.\label{n_condtion}
\eeq
By assigning the $\bfn_1,\bfn_2$ satisfying (\ref{n_condtion}), one can construct curved origami structures using (\ref{theorem3_y0}) and Table \ref{formulas0}. 
\end{corollary}
\begin{corollary}
\label{cylindrical normal}
By direct calculation from (\ref{surf1_surf2}), one can get that $\bft_\sigma\parallel\bfn_\sigma\times\bfn_\sigma',\sigma=1,2$. Assume $\bfn_\sigma'\neq0$. Then $\bft_\sigma$ can be given by
\beq
\bft_\sigma=\bfn_\sigma\times\frac{\bfn_\sigma'}{|\bfn_\sigma'|},\quad\sigma=1,2.
\eeq
If $\bfn_\sigma$ is a planar vector, $\bft_\sigma$ will be constant, which is the normal vector of the plane on which $\bfn_\sigma$ is located.

\end{corollary}

Construction of more complicated structures then involves fitting these two-surface structures together, and the group 
orbit procedure described below is a suitable method for this purpose. 

\section{Curved origami design by the group orbit procedure}
\label{group_orbit}
In this section we use a group orbit procedure to design complex origami structures, that is, origami structures are obtained by repeated application of a Euclidean group to an origami unit cell.
A key idea is that elements of a Euclidean group preserve isometries. A second key idea is that Abelian Euclidean groups preserve the matching at creases.  That is, matching of tiles at a few creases implies matching of tiles at all creases in the extended structure.

In this work, we will apply Abelian isometry groups that have been studied for piecewise 
linear origami in \cite{liu2021origami,feng2020helical,liu2022origami,feng2019phase}. The main difference between our present work and the earlier work is that in many cases our unit cells below contain interior creases, while in the earlier work the unit cells were typically single tiles.  The presence of these 
additional creases gives us extra freedom 
that we exploit to describe a continuous
folding path from a flat sheet to the folded structure.

A group element of a Euclidean group is written $g = (\bfQ|\bfc)$, where $\bfQ \in$ O(3)$= \{\bfQ : \bfQ^{\rm T} \bfQ  = \bfI \}$ and $\bfc\in\R^3$.  The action of a group element on points $\bfx \in \R^3$ is given by  $g(\bfx) = \bfQ\bfx + \bfc$.  Below, we use
the terminology ``isometry'' for the
group elements $g = (\bfQ|\bfc)$ or this action.  The subgroup of O(3) consisting
of rotations is SO(3) $= \{\bfR \in {\rm O}(3): \det \bfR = 1 \}$. 
We follow this notation below: $\bfQ$ represents a typical element in O(3) and 
$\bfR$ represents an element in SO(3).
The multiplication rule for isometries is based on the composition of mappings using the action above, i.e.,
$g_1(g_2(\bfx)) = g_1 g_2(\bfx)$ for all
$\bfx \in \R^3$, and therefore is given by
 \beq
g_1g_2=(\bfQ_1|\bfc_1)(\bfQ_2|\bfc_2)=(\bfQ_1\bfQ_2|\bfc_1+\bfQ_1\bfc_2).
\eeq
The identity is  $(\bfI|\bf0)$.

While there are many Abelian isometry groups that we could use
(see e.g., the International Tables of Crystallography \cite{Hahn2003}), we will focus on helical groups, circle groups and translation groups. We will later generalize these results to the conformal Euclidean groups, which involve also dilatations and a suitable choice of product. 

The main features of the unit cell for curved tile origami are: 1) as mentioned above, in most examples, we
introduce an extra crease in the unit cell to gain some additional freedom; 2) we often use the rulings themselves as creases,
e.g., $\overline{\bfx_a\bfx_b}$ and 
$\overline{\bfx_d\bfx_c}$ in Figure \ref{helical}(a).   The latter simplifies constructions: 3) since rulings are straight, we only have to be sure that two rulings related by a group element are the same length; 4) regardless of the two isometric mappings
meeting at a straight crease, the deformed structure
will be compatible.

Following the choice of action above,  we apply the group element $g=(\bfQ|\bfc)$ to a suitable unit cell $\Omega$ in the following way: $g(\Omega)=\bfQ\Omega+\bfc$. After this group action, each unit cell with curved tiles should fit perfectly together with its neighbors. The beauty of Abelian groups is that, if this fitting is done only for the generators of the group, then the whole structure fits together perfectly.  In short, the group builds the structure for you.  Finally, for the groups involving 
global compatibility, such as the helical and circle groups, these groups should be discrete (\cite{liu2021origami,feng2020helical} and below), which will generate discrete and globally compatible curved origami.

Extending the notation used above, we use $\Omega$ and $\calS$,
respectively, to denote the unit cell before and after folding. The origami unit cell we choose will typically contain one inner curved crease $\bfx_1(s) \subset \Omega$ and two boundary creases 
$\bfx_0(s),\ \bfx_2(s) \subset \Omega$, all parameterized by
$s\in(s_1,s_2)$, as shown in Figures \ref{helical}-\ref{conformal}(a).  The isometric images of the three curves are $\bfy_1(s)$, $\bfy_0(s)$ and $\bfy_2(s)$, respectively. Any two points on adjacent creases with the same $s$ are connected by a ruling. A ruling between $\bfy_0(s)$ and $\bfy_1(s)$ has tangent $\bft_1(s)$, and the ruling between $\bfy_1(s)$ and $\bfy_2(s)$ has tangent  $\bft_2(s)$. The deformation gradients of $\bft_1(s)$ and $\bft_2(s)$ are $\bfF_1(s)$ and $\bfF_2(s)$ respectively. The binormal vectors of the creases at $\bfy_i(s)$ are denoted by $\bfb_i(s),i=0,1,2$. $\bfP_i(s)=\bfI-2\bfb_i(s)\otimes\bfb_i(s)$ is a reflection tensor relating to the $i^{\rm th}$ crease. We use $\bfx_i,\,i=a,b,c,d$ to denote the four corner points of the reference unit cell and denote their images by $\bfy_i,\,i=a,b,c,d$, see Figures \ref{helical}-\ref{conformal}(a) and (c).

The general result we use repeatedly below is the following,
described in the context of discrete Abelian isometry groups with two generators as in Figure \ref{helical}.
Let $\calT$ be the group applied to $\Omega$ with generators  $t_1$ and $t_2$, and let $\calG$ be the group applied to $\calS$ with generators of $g_1$ and $g_2$. Let $\bfx_3(s)=\overline{\bfx_c\bfx_d},\,\bfx_4(s)=\overline{\bfx_b\bfx_a}$ denote the remaining two sides of  $\partial \Omega$ apart from $\bfx_0(s)$ and $\bfx_2(s)$. Suppose there
is an isometric mapping $\bfy: \Omega \to \calS$ satisfying $\bfy(\partial \Omega) = \partial \cal S$, that is, $\bfy(\bfx_i(s)) = \bfy_i(s), \ i = 0, 2, 3, 4$ (Figure \ref{helical}(c)).  Assume that, by adjusting the group parameters, we also arrange that 
\beqs
t_1(\bfx_0(s)) = \bfx_2(s),&& t_2(\bfx_3(s)) = \bfx_4(s),\\
g_1(\bfy_0(s)) = \bfy_2(s),&& g_2(\bfy_3(s)) = \bfy_4(s).
\eeqs
Now we apply the group $\calT$, not
just to $\partial \Omega$, but to all
of $\Omega$.  By construction, by
simply matching on  two boundaries, we achieve that 
$\calT(\Omega)$ is a perfect lattice of translated copies of $\Omega$ without
gaps (Figure \ref{helical}(b)).
Similarly, $\calG(\calS)$ is a
perfect helical structure isometrically mapped from
$\calT(\Omega)$ (Figure \ref{helical}(d)).  If the group
$\calG$ is discrete, the structure closes perfectly. That is, referring
to Figure \ref{helical}(d),  discreteness 
for a helical group has
the geometric interpretation that
the helical structure can be produced
by a ``rolling up'' construction,
and, once rolled up, there is no
seam. Equivalently, from a group
theory perspective, discreteness
means that the group acting on any
point in $\R^3$ produces a family
of points without accumulation
points. 
\subsection{Helical groups}
\label{helical_group}
\begin{figure}[ht]
\centering
\includegraphics[scale=1]{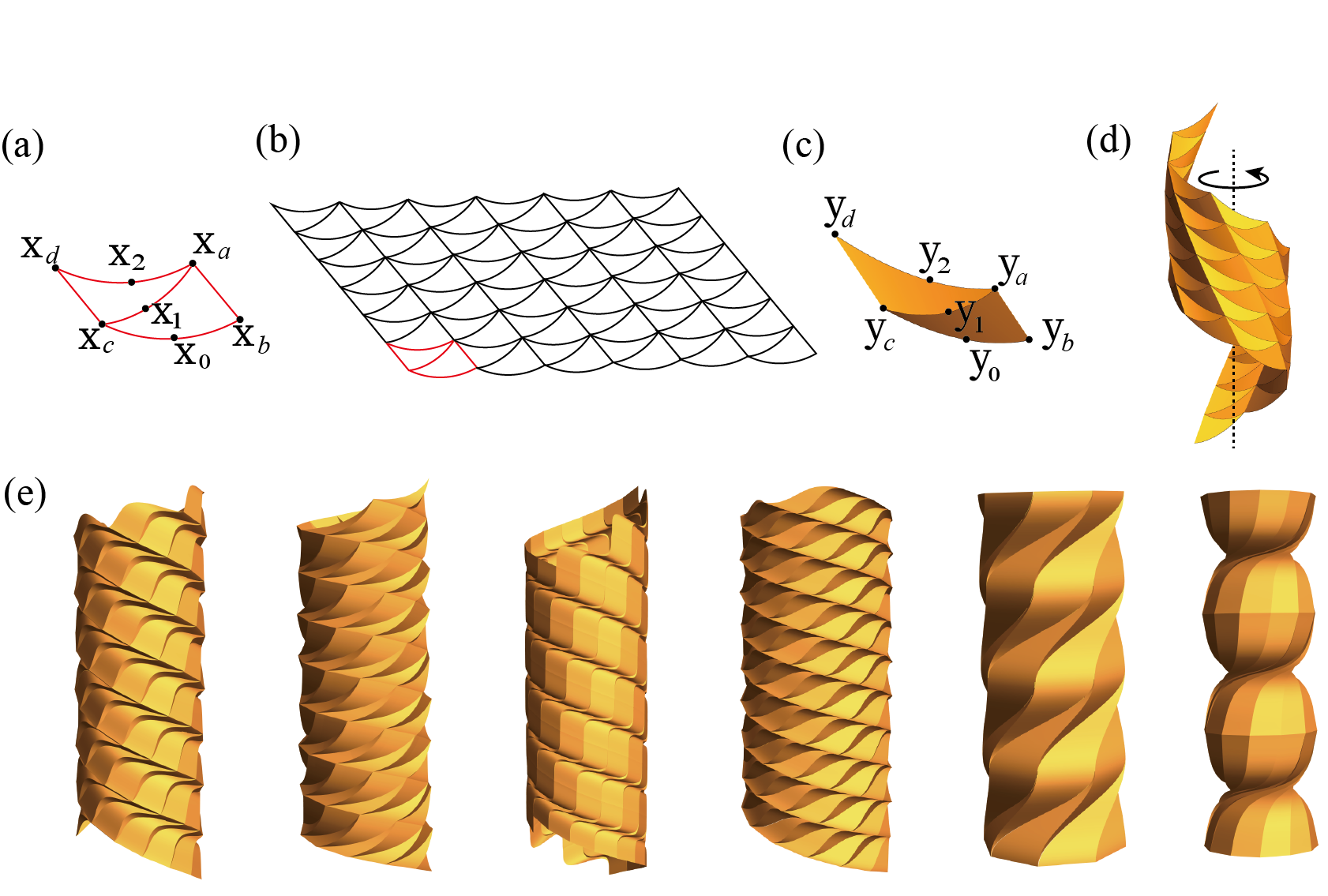}
\caption{Curved tile origami generated by helical groups. (a) Reference unit cell. (b) Reference configuration. Note that (b)
is flat sheet: the apparent curvature is an
optical illusion.  (c) Deformed unit cell. (d) Deformed configuration. (e) Some examples.}
\label{helical}
\end{figure}

Helical origami is obtained by applying a helical group to a partially folded unit cell $\calS$ as shown in Figure \ref{helical}. We
consider helical groups with two generators 
\beq
\calG(\calS)=\{g_1^pg_2^q(\calS):(p,q)\in\Z^2\},
\eeq
where $g_1,g_2$ are two screw isometries
\beq
g_1=({\bfR}_1|\tau_1\bfe_{_R}+(\bfI-\bfR_1)\bfz),\quad g_2=(\bfR_2|\tau_2\bfe_{_R}+(\bfI-\bfR_2)\bfz),
\eeq
with $\bfR_1,\bfR_2 \in$ SO(3), $\bfR_1 \ne \bfI$, $\bfe_{_R},\bfz\in\R^3, |\bfe_{_R}|=1$, $\tau_1,\tau_2\in\R$,  $\bfz\cdot\bfe_{_R}=0$ and $\bfR_1\bfe_{_R}=\bfR_2\bfe_{_R}=\bfe_{_R}$. Let $\theta_1$ and $\theta_2$ denote the rotation angles of $\bfR_1$ and $\bfR_2$,  respectively. These parameters are subject to discreteness conditions 
\beq
p^\star\tau_1+q^\star\tau_2=0,\quad p^\star\theta_1+q^\star\theta_2=2\pi,\label{glob_comp}
\eeq 
where $(p^\star,q^\star)\in\Z^2$ is some pair of integers \cite{feng2019phase}.  

In the reference domain, the reference unit cell $\Omega$ shown in Figure \ref{helical}(a) consists of two triangle-like flat regions that meet at the curved crease $\bfx_1(s), s_1<s<s_2$. The two boundary creases $\bfx_0(s)$ and $\bfx_2(s)$ differ by a translation. All creases satisfy the
smoothness conditions of Theorem \ref{theorem2}. The other two straight boundaries on the left- and right-handed sides are chosen to be two rulings after deformation. The overall reference domain can be obtained by a translation group
\beq
\calT(\Omega)=\{t_1^pt_2^q(\Omega):(p,q)\in\Z^2\},
\eeq
where 
\beq
t_1=({\bfI}|{\bfc_1}),\quad t_2=(\bfI|\bfc_2),
\eeq
with $\bfc_1,\bfc_2\in\R^3$ and $t_1^pt_2^q(\Omega)=\Omega+p\bfc_1+q\bfc_2$, $p,q \in \Z$. See Figure \ref{helical}(b), which is a curved-crease generalization of the Kresling pattern.

For a suitable unit cell, we apply $\calT$ to its reference configuration $\Omega$ and apply $\calG$ to its deformed configuration $\calS$. The curved creases between adjacent unit cells should be compatible before and after folding. Then, at the curved crease, we have the local compatibility conditions 
\beqs
\bfx_2(s)&=&t_1(\bfx_0(s))=\bfx_0(s) +\bfc_1,\\
\bfy_2(s)&=&g_1(\bfy_0(s))=\bfR_1\bfy_0(s)+\tau_1\bfe_{_R}+(\bfI-\bfR_1)\bfz.\label{sec_derivatives}
\eeqs
Thus, $\bfx_2'=\bfx_0',\; \bfy_2'=\bfR_1\bfy_0'$.
Combining with $\bfy_0'=\bfF_1\bfx_0'$ and $\bfy_2'=\bfF_2\bfx_2'$, we have $\bfy_2'=\bfR_1\bfF_1\bfx_2'$,
where $\bfF_2$ and $\bfR_1\bfF_1$ correspond to the deformation gradients of the surfaces on opposite sides of $\bfy_2$. From Corollary \ref{normal_grad}, the deformation gradients satisfy $\bfF_2=\bfP_1\bfF_1$ and $\bfF_2=\bfP_2\bfR_1\bfF_1$. So 
\beq
\bfP_2\bfP_1=\bfR_1.\label{p2p1R1}
\eeq 
Since $\bfR_1$ is constant, we also have $\bfy_2''=\bfR_1\bfy_0''$ from (\ref{sec_derivatives}). Replacing $\bfR_1$ by $\bfP_2\bfP_1$ we have 
\beq
\bfy_2'=\bfP_2\bfP_1\bfy_0'=\bfP_1\bfy_0',\quad \bfy_2''=\bfP_2\bfP_1\bfy_0''=\bfP_1\bfy_0''.\label{y2_p1}
\eeq
Because $\bfy_2''=(\bfP_1\bfy_0')'=\bfP_1\bfy_0''+\bfP_1'\bfy_0'$, we get $\bfP_1'\bfy_0'=0$, that is, $(\bfb_1\cdot\bfy_0')\bfb_1'+(\bfb_1'\cdot\bfy_0')\bfb_1=0$. By simple argument, one can see $\bfb_1$ is constant. In the same way, we can conclude that
$\bfb_0,\bfb_2$ are constant, which implies the creases $\bfy_0(s),\bfy_1(s)$, and $\bfy_2(s)$ are planar curves. According to Corollary \ref{planar_crease} for planar creases, we get $\bft_1=-\bfP_1\bft_2$ and $\bfe_1=-\bfe_2$. Since the rulings in the whole reference domain are periodic, the reference rulings will be parallel, which gives cylindrical surfaces for the deformed unit cell.
Of course, as can be seen from Figure \ref{helical}, different unit cells have
different cylindrical surfaces. We want
to emphasize that the procedure outlined
here gives one simple way to make helical origami structures with curved tiles, but the more general procedure described at the beginning of this section gives other possibilities.

In the implementation, if the planar crease $\bfy_1(s)$ and the constant ruling $\bft_1$ of one tile are given, the ruling of the other tile will be found by $\bft_2=-\bfP_1\bft_1$. Since $\bfy_0(s)$ is a planar curve in the tile, a simple way to get $\bfy_0(s)$ is to cut off the rulings with a plane, whose normal vector will be the binormal $\bfb_0$ of $\bfy_0(s)$. Once $\bfy_0$ is found, $\bfy_2$ will be determined since $\bfb_2=-\bfP_1\bfb_0$ from (\ref{y2_p1}). Specifically, the two boundaries
$\bfy_0(s)$ and $\bfy_2(s)$ can be obtained by
\beq
\bfy_0=\bfy_1-\frac{(\bfy_1-\bfy_1(s_1))\cdot\bfb_0}{\bft_1\cdot\bfb_0}\bft_1,\quad\bfy_2=\bfy_1+\frac{(\bfy_1-\bfy_1(s_2))\cdot\bfb_0}{\bft_1\cdot\bfb_0}\bft_2.\label{y0-y2}
\eeq
The creases $\bfy_0$ and $\bfy_2$ given by (\ref{y0-y2}) have the same configuration since (\ref{y2_p1}) is satisfied. Also, $\bfx_0'=\bfx_2'$ will be automatically satisfied in the reference domain, which can be verified by using $\bfx_0'=\bfF_1^{\rm T}\bfy_0'$ and $\bfx_2'=\bfF_2^{\rm T}\bfy_2'$ and Corollary \ref{planar_crease}. Therefore, the adjacent unit cells will be compatible at curved creases in both reference and deformed domains after group operations. An algorithm to construct the helical curved origami
follows.
\begin{breakablealgorithm}
\caption{Helical group} 
\hspace*{0.02in} {\bf Input:} 
planar crease $\bfy_1(s)$, constant binormal $\bfb_0$, constant ruling $\bft_1$ of one surface.\\
\hspace*{0.02in} {\bf Output:} 
helical curved origami.\\
\hspace*{0.02in} {Steps:}
\begin{algorithmic}[1]
\State {Find the unit cell $\calS$:
\begin{itemize}
\item Find the deformed rulings of the second surface:
\beq
\bft_2=-\bfP_1\bft_1.
\eeq
    \item Find the two boundary creases $\bfy_0(s)$ and $\bfy_2(s)$:
\beqs
\bfy_0(s)&=&\bfy_1(s)-\frac{(\bfy_1(s)-\bfy_1(s_1))\cdot\bfb_0}{\bft_1\cdot\bfb_0}\bft_1,\\
\bfy_2(s)&=&\bfy_1(s)+\frac{(\bfy_1(s)-\bfy_1(s_2))\cdot\bfb_0}{\bft_1\cdot\bfb_0}\bft_2,
\eeqs
\item Construct the deformed unit cell:
\beqs
\calS=\left\{\begin{array}{ll}
(1-v)\bfy_1(s)+v\bfy_0(s)\\
(1-v)\bfy_1(s)+v\bfy_2(s)
\end{array}:0<v<1,\,s_1<s<s_2\right\}
\eeqs
\end{itemize} 
}
\State{Find the group parameters of $\calG$:
\beqs
\bfe_{_R}&=&\frac{\bfb_0\times\bfb_1}{|\bfb_0\times\bfb_1|}\label{helical_axis}\\
\theta_1&=&\text{sign}(\bfe_{_R}\cdot(\bfy_a-\bfy_b)\times(\bfy_d-\bfy_c))\arccos\left(\frac{\bfA(\bfy_a-\bfy_b)\cdot\bfA(\bfy_d-\bfy_c)}{|\bfy_a-\bfy_b||\bfy_d-\bfy_c|}\right)\\
\theta_2&=&\text{sign}(\bfe_{_R}\cdot(\bfy_a-\bfy_d)\times(\bfy_b-\bfy_c))\arccos\left(\frac{\bfA(\bfy_a-\bfy_d)\cdot\bfA(\bfy_b-\bfy_c)}{|\bfy_a-\bfy_d||\bfy_b-\bfy_c|}\right)\\
\tau_1&=&\bfe_{_R}\cdot(\bfy_a-\bfy_d)\\
\tau_2&=&\bfe_{_R}\cdot(\bfy_a-\bfy_b)\\
\bfz&=&\bar\bfR_2\bfA(\bfy_a-\bfR_2\bfy_d)\label{helical_z}
\eeqs
where $\bfA=\bfI-\bfe_{_R}\otimes\bfe_{_R}$, $\bar\bfR_2=(\bfI+\bfe_{_R}\otimes\bfe_{_R}-\bfR_2)^{-1}-\bfe_{_R}\otimes\bfe_{_R}$, $\bfy_a=\bfy_2(s_1),\bfy_b=\bfy_0(s_1)$, and $\bfy_c=\bfy_0(s_2),\bfy_d=\bfy_2(s_2)$.}
\State{Apply the helical group to the unit cell: \beq\calG(\calS)=\{g_1^pg_2^q(\calS):(p,q)\in\Z^2\}.\eeq}
\end{algorithmic}
\end{breakablealgorithm}
Here if $\bfy_1(s)$ is fixed, the group parameters (\ref{helical_axis}-\ref{helical_z}) will rely on $\bfb_0$ and $\bft_1$.  In this algorithm, to get a globally compatible helical origami, suitable $\bfb_0$ and $\bft_1$ are chosen in advance by numerically solving (\ref{glob_comp}) with certain $(p^\star,q^\star)$ by plugging in the group parameters with respect to $\bfb_0$ and $\bft$. See some helical curved tile origami in Figure \ref{helical}.
\subsection{Circle groups}
\label{circ1}
\begin{figure}[ht]
\centering
\includegraphics[scale=1]{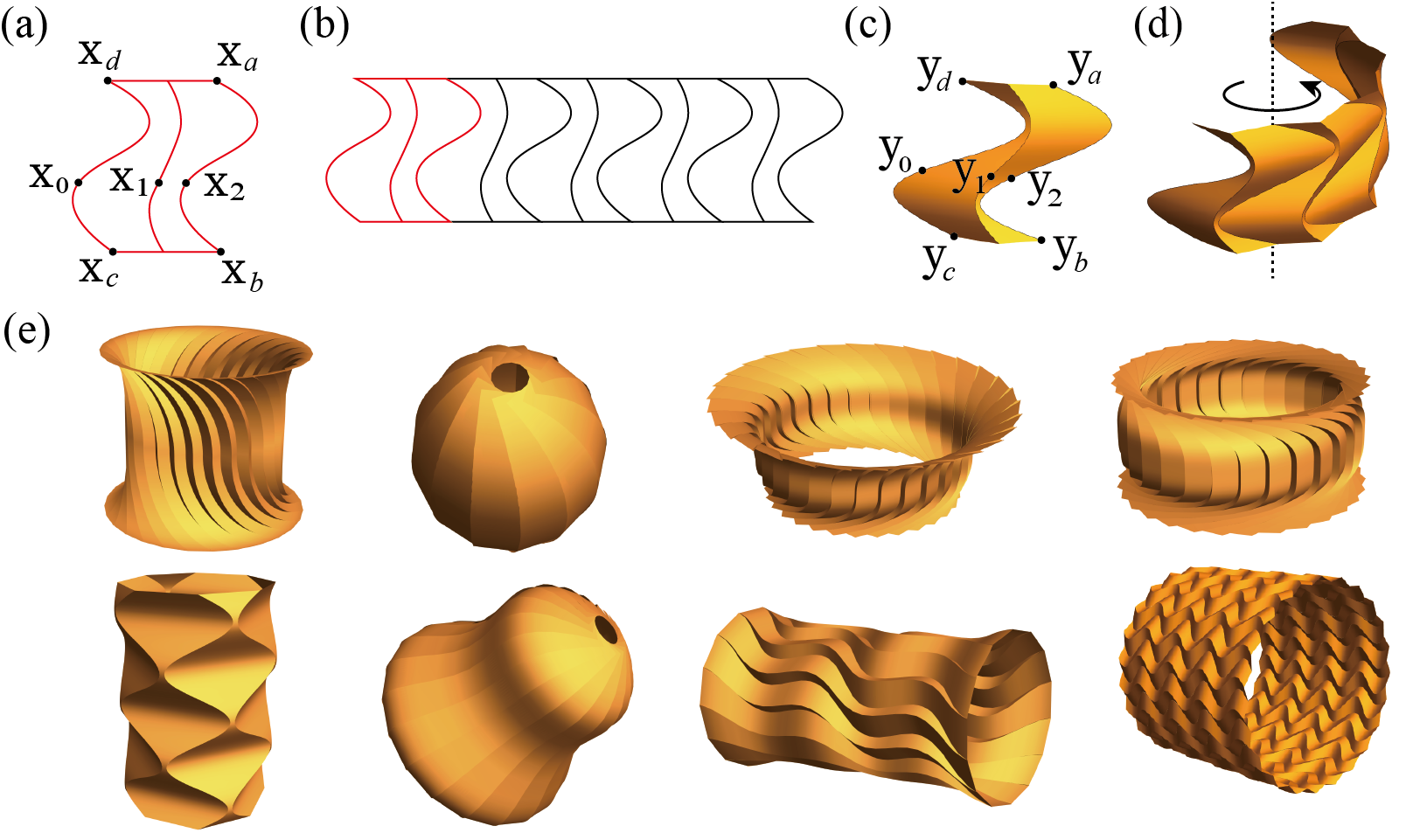}
\caption{Curved tile origami generated by circle groups. (a) Reference unit cell. (b) Reference configuration. (c) Deformed unit cell. (d) Deformed configuration. (e) Some examples.}
\label{rotation}
\end{figure}

In this subsection, we study the degenerate case where a circle group is applied to a partially folded unit cell as shown in Figure \ref{rotation}. The reference unit cell in this case consists of two parallelogram-like planar sheets.  The whole reference domain can be obtained by taking the product of a translation to $\Omega$. To define this translation, let
\beq
t=(\bfI|\bfc)\label{case1_ref}.
\eeq
Then the overall flat tessellation in Figure \ref{rotation}(b) is obtained by 
\beq
\calT(\Omega)=\{t^p(\Omega),\,p\in\Z\}.
\eeq
For the deformed domain define
\beq
g=(\bfR|(\bfI-\bfR)\bfz), \label{case1_def}
\eeq
where $\bfR$ is a rotation of angle $2\pi/n$ with $n\in\Z$, $\bfz\in\R^3$ and $\bfz\cdot\bfe_{_R}=0$, $\bfe_{_R}$ is the unit vector along the axis of $\bfR$. Then the overall deformed configuration is obtained by
\beq
\calG(\calS)=\{g^p(\calS),\,p\in\Z\}.
\eeq 
By following the same steps in the helical origami, one can get the same conclusion that all the creases are planar curves and all surfaces are cylindrical surfaces. The algorithm to construct curved origami using a circle group follows. 
\begin{breakablealgorithm}
\label{trans_circle}
\caption{Circle group} 
\hspace*{0.02in} {\bf Input:} 
planar crease $\bfy_1(s)$ and constant ruling $\bft_1$ of one surface, the order of group $\calG$: $n$.\\
\hspace*{0.02in} {\bf Output:} curved origami with rotational symmetry.\\
\hspace*{0.02in} {Steps:}
\begin{algorithmic}[1]
\State{Find the unit cell:\begin{itemize}
\item Find the deformed rulings of the second surface:
\beq
\bft_2=-\bfP_1\bft_1
\eeq
    \item Find the axis of $\bfR$ and the two boundary creases $\bfy_0(s)$ and $\bfy_2(s)$:
\beqs
\bfe_{_R}&=&\frac{\bft_1\times\bft_2}{|\bft_1\times\bft_2|},\\
\bfy_0(s)&=&(\bfI-\frac{\bft_1\otimes\bfb_0}{\bft_1\cdot\bfb_0})\bfy_1(s)+a\bft_1,\quad a\in\R\\
\bfy_2(s)&=&(\bfI-\frac{\bft_2\otimes\bfb_2}{\bft_2\cdot\bfb_2})\bfy_1(s)+b\bft_2,\quad b\in\R
\eeqs
where $\bfb_0=\bfR_{\alpha}\bfb_1,\bfb_2=-\bfP_1\bfb_0$
and $\bfR_\alpha$ is a rotation about axis $\bfe_{_R}$ with angle $\alpha=\frac{\pi}{2}-\frac{\pi}{n},n\in\Z$. 
\item Construct the deformed unit cell:
\beqs
\calS=\left\{\begin{array}{ll}
(1-v)\bfy_1(s)+v\bfy_0(s)\\
(1-v)\bfy_1(s)+v\bfy_2(s)
\end{array}:0<v<1,\,s_1<s<s_2\right\}
\eeqs
\end{itemize} 
}
\State{Find the group parameters of $\calG$:
\beq
\bfz=\bar\bfR\bfA(\bfy_a-\bfR\bfy_d).
\eeq
where $\bfA=\bfI-\bfe_{_R}\otimes\bfe_{_R},\ \bar\bfR=(\bfI+\bfe_{_R}\otimes\bfe_{_R}-\bfR)^{-1}-\bfe_{_R}\otimes\bfe_{_R}$.}
\State{Apply the circle group to the unit cell: \beq
\calG(\calS)=\{g^p(\calS),\,p\in\Z\}.\eeq}
\end{algorithmic}
\end{breakablealgorithm}
\subsection{Translation groups}
\begin{figure}[ht]
\centering
\includegraphics[scale=1]{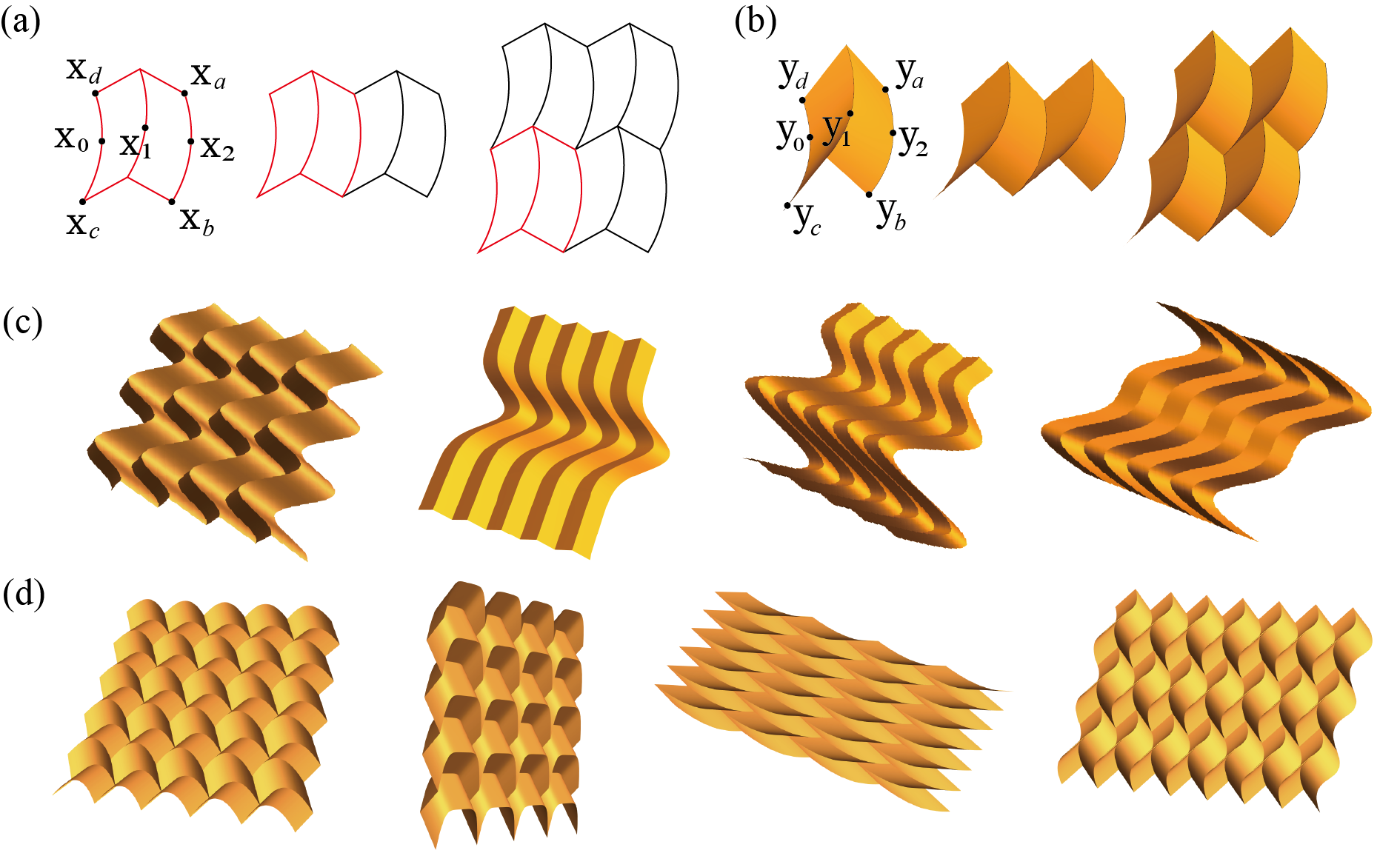}
\caption{Curved tile origami generated by translation groups. (a) Reference configurations (flat). (b) Deformed configurations. (c) Some examples generated by the translation groups with one generator. (d) Some examples generated by the translation groups with two generators.}
\label{translation}
\end{figure}

Periodic curved origami is obtained by applying translation groups to the unit cell $\calS$. The schematics of the reference and deformed unit cells are shown in Figure \ref{translation}. 
In the reference configuration, the crease $\bfx_2(s)$ is obtained by translating $\bfx_0(s)$ by a translation $\bfc_1$. Similarly, in the deformed configuration, $\bfy_2(s)$ is obtained from $\bfy_0(s)$ by $\bfc_2$. Define
\beq
t_1=(\bfI|\bfc_1),\quad g_1=(\bfI|\bfc_2),
\eeq
then the whole reference domain and deformed domain are found by
\beq
\calT(\Omega)=\{t_1^p(\Omega),\,p\in\Z\},\quad\calG(\calS)=\{g_1^p(\calS),\,p\in\Z\}.
\eeq
Since $\bfx_2=t_1(\bfx_0)=\bfx_0+\bfc_1$ and $\bfy_2=g_1(\bfy_0)=\bfy_0+\bfc_2$, we have $\bfx_0'=\bfx_2'$ and $\bfy_0'=\bfy_2'$. Thus, $\bfy_2'=\bfF_2\bfx_2'=\bfy_0'=\bfF_1\bfx_0'=\bfF_1\bfx_2'$. At crease $\bfy_1$, the compatibility condition gives $\bfF_1\bfx_1'=\bfF_2\bfx_1'$. So 
\beq
(\bfF_1-\bfF_2)\bfx_1'=0,\quad (\bfF_1-\bfF_2)\bfx_2'=0 \quad {\rm with }\;\bfF_1 \ne \bfF_2.
\eeq
We get $\bfx_2'=c\bfx_1',c\in\R$. Therefore, $\bfy_2'=\bfF_2\bfx_2'=c\bfF_2\bfx_1'=c\bfy_1'$. Similarly, we have $\bfy_1'=c\bfy_0'$. Then $\bfy_2'=\bfy_0'$ implies $c=1$. Thus,
\beq
\bfy_0'=\bfy_1'=\bfy_2'.\label{trans_y012}
\eeq
Let $\bfy_2=\bfy_1+ v_1\bft_1$ and $\bfy_2=\bfy_1+v_2\bft_2$. Then $v_1\bft_1$ and $v_2\bft_2$ will be constant independent of $s$ from (\ref{trans_y012}). Therefore, the two surfaces in $\calS$ will be cylindrical surfaces, but the deformed creases are not necessarily planar curves in this case.

In the implementation, we get the cylindrical surfaces by assigning planar normal vectors $\bfn_1$ and $\bfn_2$ according to Corollary \ref{theorem3} and Corollary \ref{cylindrical normal}. The algorithm is given below.
\begin{breakablealgorithm}
\caption{Translation group} 
\hspace*{0.02in} {\bf Input:} 
the two planar normal vectors $\bfn_1(s),\bfn_2(s)$.\\
\hspace*{0.02in} {\bf Output:} periodic curved origami.\\
\hspace*{0.02in} {Steps:}
\begin{algorithmic}[1]
\State{Find the unit cell:\begin{itemize}
\item Find the crease $\bfy_1(s)$:
\beq
\bfy_1(s)=\bar\bfy_1+\int_{s_1}^s\frac{\bfn_1\times\bfn_2}{|\bfn_1\times\bfn_2|}dr.
\label{y1_cylindrical}
\eeq
where $\bar\bfy_1=\bfy_1(s_1)$ is a given constant.
\item Find the two constant rulings $\bft_1$ and $\bft_2$:
\beq
\bft_1=\bfn_1\times\frac{\bfn_1'}{|\bfn_1'|},\quad\bft_2=\bfn_2\times\frac{\bfn_2'}{|\bfn_2'|}.
\eeq
    \item Find the two boundary creases $\bfy_0(s)$ and $\bfy_2(s)$:
\beqs
\bfy_0(s)&=&\bfy_1(s)+v_1\bft_1,\\
\bfy_2(s)&=&\bfy_1(s)+v_2\bft_2,
\eeqs
where $v_1,v_2>0$ are constant.
\item Construct the deformed unit cell:
\beqs
\calS=\left\{\begin{array}{ll}
(1-v)\bfy_1(s)+v\bfy_0(s)\\
(1-v)\bfy_1(s)+v\bfy_2(s)
\end{array}:0<v<1,\,s_1<s<s_2\right\}.
\eeqs
\end{itemize} 
}
\State{Find the group parameters of $\calG$:
\beq
\bfc_2=\bfy_a-\bfy_d.
\eeq}
\State{Apply the circle group to the unit cell: \beq
\calG(\calS)=\{g^p(\calS),\,p\in\Z\}.\eeq}
\end{algorithmic}
\end{breakablealgorithm}
Since the upper boundary $\widehat{\bfx_d\bfx_a}$ is a translation of  the lower boundary $\widehat{\bfx_c\bfx_b}$ in $\Omega$, we can also introduce the second translation generator to $\Omega$: $t_2=(\bfI|\bfc_3)$, where $\bfc_3=\bfx_a-\bfx_b$. Then the reference domain is given by
\beq
\calT(\Omega)=\{t_1^pt_2^q(\Omega),p,q\in\Z\},
\eeq
see Figure \ref{translation}(a)$_3$. In the deformed domain, $\bfy(g_3(\Omega))$ is automatically compatible with $\bfy(\Omega)$. Let $\bfc_4=\bfy_a-\bfy_b$ and $g_2=(\bfI|\bfc_4)$, the whole deformed configuration is found by
\beq
\calG(\calS)=\{g_1^pg_2^q(\calS),(p,q)\in\Z^2\}.
\eeq
See examples in Figure \ref{translation}(d).

\subsection{More on circle groups}
\label{circ2}
\begin{figure}[ht]
\centering
\includegraphics[scale=1]{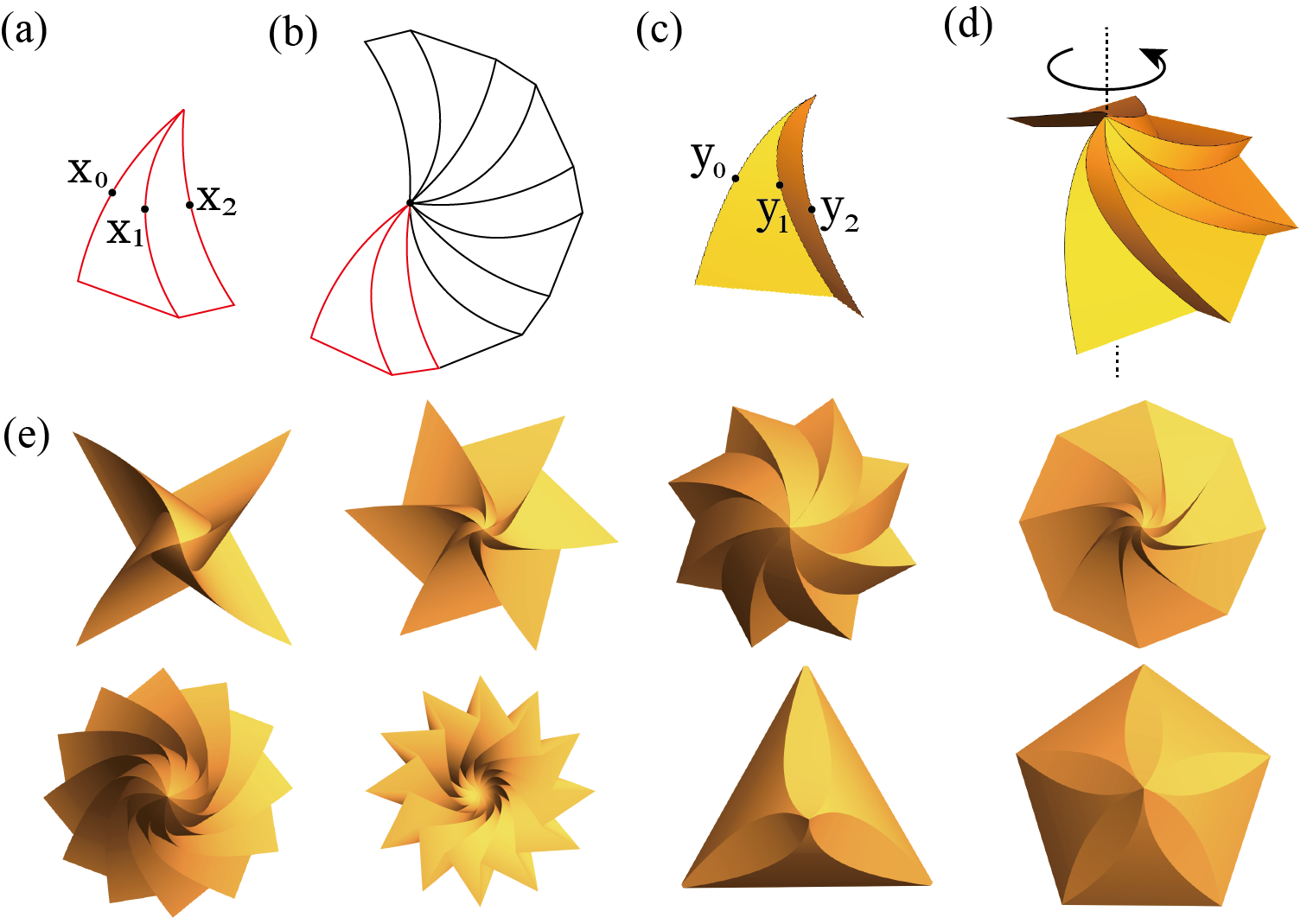}
\caption{Curved tile origami generated by circle groups. (a) Reference unit cell. (b) Reference configuration. (c) Deformed unit cell. (d) Deformed configuration. (e) Some examples.}
\label{rotats}
\end{figure}

As an alternative to the constructions of
Section \ref{circ1}, in this subsection we apply circle groups to both the reference unit cell and the deformed unit cell, see Figure \ref{rotats}. Define
\beq
t=({\bfR}_1|{\bf0}),\quad g=(\bfR_2|\bf0),
\eeq
where $\bfR_1\in$ SO(2) and $\bfR_2\in$ SO(3). The reference and deformed domains are given by
\beq
\calT(\Omega)=\{t^p(\Omega):p\in\Z\},\quad\calG(\calS)=\{g^p(\calS):p\in\Z\}.
\eeq
Let the normal vectors of $\calS_1$ and $\calS_2$ on the both sides of $\bfy_1(s)$ be $\bfn_1(s)$ and $\bfn_2(s)$. Then the normal vectors on the both sides of $\bfy_2(s)$ will be $\bfn_2(s)$ and $\bfR_2\bfn_1(s)$. Let $\bfy_1(s_1)=\bfy_2(s_1)=0$. According to Corollary \ref{theorem3}, an algorithm is presented in Algorithm \ref{circle_groups}.

\begin{breakablealgorithm}
\label{circle_groups}
\caption{Circle groups}
\hspace*{0.02in} {\bf Input:} 
the normal vectors $\bfn_1$ and $\bfn_2$, rotation $\bfR_2$.\\
\hspace*{0.02in} {\bf Output:} periodic curved origami.\\
\hspace*{0.02in} {Steps:}
\begin{algorithmic}[1]
\State{Find the unit cell:\begin{itemize}
\item Find the crease $\bfy_1(s)$:
\beq
\bfy_1(s)=\int_{s_1}^s\frac{\bfn_1\times\bfn_2}{|\bfn_1\times\bfn_2|}dr.
\eeq
\item Find the two boundary creases $\bfy_0(s)$ and $\bfy_2(s)$:
\beqs
\bfy_2(s)&=&\int_{s_1}^s\frac{\bfR_2\bfn_1\times\bfn_2}{|\bfR_2\bfn_1\times\bfn_2|}dr,\\
\bfy_0(s)&=&\bfR_2^{\rm T}\bfy_2(s).
\eeqs
\item Construct the deformed unit cell:
\beqs
\calS=\left\{\begin{array}{ll}
(1-v)\bfy_1(s)+v\bfy_0(s)\\
(1-v)\bfy_1(s)+v\bfy_2(s)
\end{array}:0<v<1,\,s_1<s<s_2\right\}.
\eeqs
\end{itemize} 
}
\State{Apply the circle group to the unit cell: \beq
\calG(\calS)=\{g^p(\calS),\,p\in\Z\}.\eeq}
\end{algorithmic}
\end{breakablealgorithm}
See some examples in Figure (\ref{rotats}).
\subsection{Conformal groups}
Conformal groups, which include dilatations as well
as rotations and translations, can be used to construct origami structures. At first, one might think that the dilatations present in 
conformal groups might be incompatible with the pure
isometric deformations possible in thin tiles.  However, one can simply recognize that the basic
scaling law of nonlinear elasticity theory,
$\bfy(\bfx,t) \to \eta \bfy((1/\eta)\bfx, (1/\eta)t)$, which preserves the deformation
gradient, stress, balance of linear momentum
(no body force), isometric deformations, etc., can be invoked by simply
allowing neighboring tiles to be related by the 
appropriate dilatation, as well as a possible rotation and
translation.  To be compatible with elasticity
scaling, the same dilatation must be used on
corresponding tiles in
the reference and deformed domains.  Conformal
groups necessarily have at least one accumulation
point, so this point must of course be
avoided, by e.g., introducing a small cut-out.

\begin{figure}[ht]
\centering
\includegraphics[scale=1]{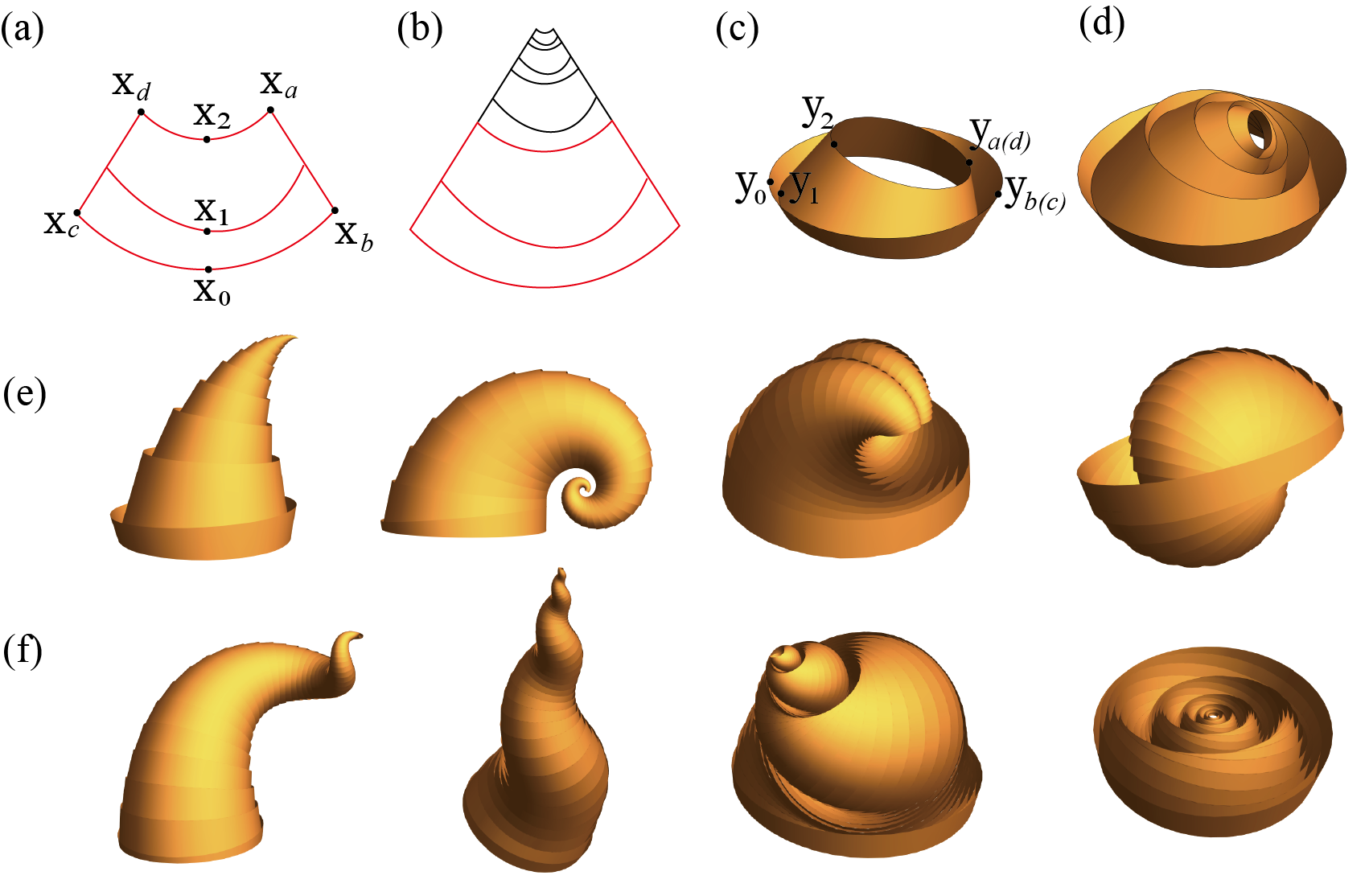}
\caption{Curved tile origami generated by conformal groups. (a) Reference unit cell. (b) Reference configuration. (c) Deformed unit cell. (d) Deformed configuration. (e-f) Some examples.}
\label{conformal}
\end{figure}

Since neighboring tiles are related by dilatation,
conformal groups feature structures that have a natural mechanism of growth, beginning from
their accumulation points (Figure \ref{conformal}). In this regard our structures here are
similar to some discussed by Thompson
\cite{thompson1942growth}, especially
the examples of sea shells and horns.
While
isometries do not play a significant 
role in his analysis, the
group orbit procedure  (described in 
words) applied with conformal groups is central to his work.

We say that $\calG$ is a conformal Euclidean group if $\calG$ is a group of affine linear mappings of the form $g=(\eta\bfQ|\bfc)$, where $\eta>0$ and $\eta\neq1$, $\bfQ\in$O(3), $\bfc\in\R^3$. We will be particularly interested in the generators of the form
\beq
g_i=(\eta_i\bfR_{\theta_i}|(\bfI-\eta_i\bfR_{\theta_i})\bfc_i),
\eeq
where $\eta_i>0$ and $\eta_i\ne1$, $\bfR_{\theta_i}\in$ SO(3) denotes a rotation of angle $\theta_i$, $\bfc_i\in\R^3$. Let the generators for the reference domain and deformed domain be $t$ and $g$ respectively, which are defined by 
\beq
t=(\eta\bfI|(1-\eta)\bfc_0),\quad g=(\eta\bfR|(\bfI-\eta\bfR)\bfc).
\eeq
The reference unit cell $\Omega$ is shown in Figure \ref{conformal}(a). The whole reference domain can be obtained by
\beq
\calT(\Omega)=\{t^p(\Omega):p\in\Z\},
\eeq
where $t^p=(\eta^p\bfI|(1-\eta^p)\bfc_0)$. The deformed unit cell $\calS$ is shown in Figure \ref{conformal}(c), and the deformed configuration in Figure \ref{conformal}(d) is obtained by 
\beq
\calG(\calS)=\{g^p(\calS):p\in\Z\},
\eeq
where $g^p=(\eta^p\bfR_{p\theta}|(\bfI-\eta^p\bfR_{p\theta})\bfc)$ and the subscript denotes the rotational angle of $\bfR$. Since $\bfx_2=t(\bfx_0)=\eta\bfx_0+(1-\eta)\bfc_0$ and $\bfy_2=g(\bfy_0)=\eta\bfR\bfy_0+(\bfI-\eta\bfR)\bfc$, we have 
\beq
\bfx_2'=\eta\bfx_0',\quad\bfy_2'=\eta\bfR\bfy_0' \label{conf_relation}
\eeq
Combining this with $\bfy_0'=\bfF_1\bfx_0'$ and $\bfy_2'=\bfF_2\bfx_2'$, we get $\bfy_2'=\bfR\bfF_1\bfx_2'=\bfF_2\bfx_2'$. 
This gives the compatibility condition at $\bfx_2(s)$. Because $\bfF_2=\bfP_1\bfF_1,\,\bfR\bfF_1=\bfP_2\bfF_2$, we have $\bfF_1=\bfP_1\bfP_2\bfR\bfF_1$. Notice that $\bfP_1\bfP_2\bfR\in$ SO(3). We get $\bfP_1\bfP_2\bfR=\bfI$ and therefore $\bfP_2\bfP_1=\bfR$. We obtain the same conclusion that the creases are planar curves as discussed in Section \ref{helical_group}. Thus, the rulings in the reference domain with same $s$ are collinear. Since $\eta\neq1$, the 
extension of all reference rulings, which can be given by $\bfx_2-\bfx_0$, will intersect at a common point because of the proportional relationship of $\bfx_2'=\eta\bfx_0'$. Therefore, the deformed surfaces are generalized conical surfaces. Algorithm 5 is used to construct conformal curved origami.
\begin{breakablealgorithm}
\caption{Conformal group} 
\hspace*{0.02in} {\bf Input:} 
two planar creases on the conical surface: $\bfy_0(s)$ and $\bfy_1(s)$.\\
\hspace*{0.02in} {\bf Output:} conformal curved origami.\\
\hspace*{0.02in} {Steps:}
\begin{algorithmic}[1]
\State{Find the unit cell:
\begin{itemize}
\item Find the deformed rulings:
\beq
\bft_1(s)=\frac{\bfy_0(s)-\bfy_1(s)}{|\bfy_0(s)-\bfy_1(s)|},\quad\bft_2(s)=-\bfP_1\bft_1(s).
\eeq
    \item Find the crease $\bfy_2(s)$:
\beq
\bfy_2(s)=\bigg(\bfI-\frac{\bft_2(s)\otimes\bfb_2}{\bft_2(s)\cdot\bfb_2}\bigg)\bfy_1(s)+b\, \bft_2(s),
\eeq
where $b\in\R$ and $\bfb_2=-\bfP_1\bfb_0$.
\item Construct the deformed unit cell:
\beqs
\calS=\left\{\begin{array}{ll}
(1-v)\bfy_1(s)+v\bfy_0(s)\\
(1-v)\bfy_1(s)+v\bfy_2(s)
\end{array}:0<v<1,\,s_1<s<s_2\right\}
\eeqs
\end{itemize} 
}
\State{Find the group parameters of $\calG$:
\beqs
\eta&=&\frac{|\bfy_2-\bfy_b|}{|\bfy_0-\bfy_a|}\label{conf_lambda}\\
\bfe_{_R}&=&\frac{\tilde\bfe_{_R}}{|\tilde\bfe_{_R}|},\quad\tilde\bfe_{_R}=(\frac{\bfy_0-\bfy_a}{|\bfy_0-\bfy_a|}-\frac{\bfy_2-\bfy_b}{|\bfy_2-\bfy_b|})\times(\frac{\bfy_0-\bfy_2}{|\bfy_0-\bfy_2|}-\frac{\bfy_a-\bfy_b}{|\bfy_a-\bfy_b|})\\
\theta&=&\text{sign}(\bfe\cdot(\bfy_0-\bfy_a)\times(\bfy_2-\bfy_b))\arccos\left(\frac{\bfA(\bfy_0-\bfy_a)\cdot\bfA(\bfy_2-\bfy_b)}{|\bfA(\bfy_0-\bfy_a)||\bfA(\bfy_2-\bfy_b)|}\right)\\
\bfc&=&(\bfI-\eta\bfR)^{-1}(-\eta\bfR\bfy_0+\bfy_2)\label{conf_c}
\eeqs
where $\bfe_{_R}$ is the axis of $\bfR$ and $\bfA=\bfI-\bfe_{_R}\otimes\bfe_{_R}$.}
\State{Apply the circle group to the unit cell: \beq
\calG(\calS)=\{g^p(\calS),\,p\in\Z\}.\eeq}
\end{algorithmic}
\end{breakablealgorithm}
Thus, applying the group $\calG$ with the group parameters listed above to the unit cell, we can get  conformal origami structures. Some examples are shown in Figure \ref{conformal}(e-f). 
\section{Curved tile generalizations of the Miura pattern}
\label{multiple-fold}

\begin{figure}[H]
\centering
\includegraphics[scale=1]{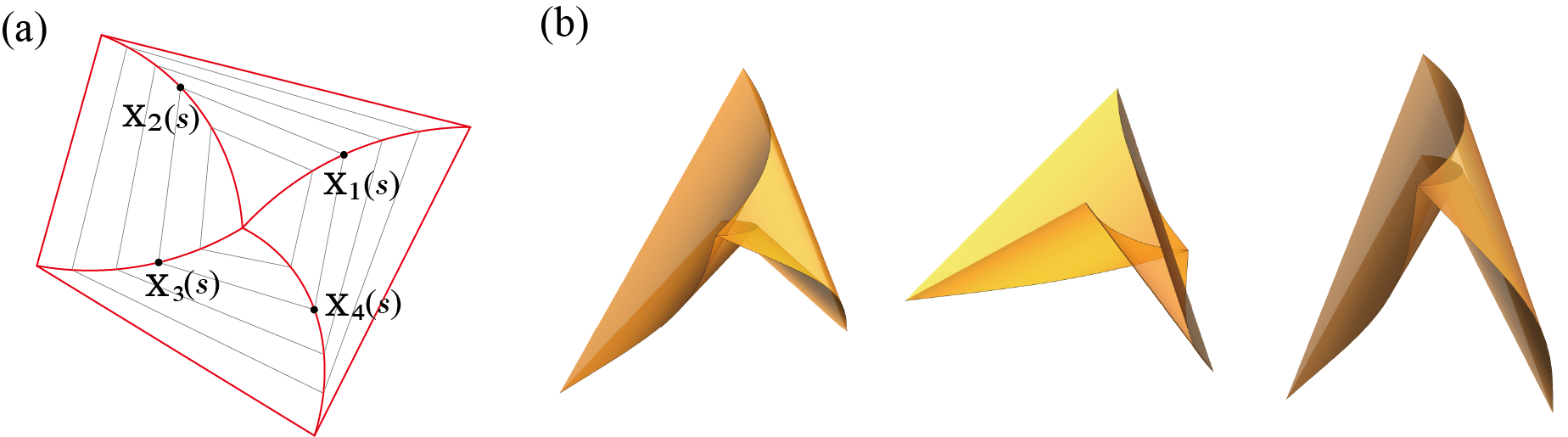}
\caption{Curved Miura origami. (a) Schematic of reference configuration, where the gray lines represent the reference rulings. (b) Some examples.}
\label{fig:four_fold}
\end{figure}

When folding a piece of paper in a
random way, it is commonly observed that several creases intersect at one point. In this section, we study the kinematics of curved tiles in this situation and discuss one special case where the rulings of adjacent tiles intersect at creases and form closed loops around the vertex, that is, all creases can be parameterized by one parameter $s$. A necessary and sufficient condition for constructing such compatible curved origami is given below.
\begin{theorem}
\label{multiple_creases}
Consider an origami structure with $n\; (n>2)$ curved creases intersecting at one vertex. Suppose the reference rulings form closed loops around the vertex in the reference domain. Denote deformed creases counterclockwise to be $\bfy_i(s),\, 0<s<s_1,\,i=1,\dots,n$. The points of $\bfy_i(s),\,i=1,\dots,n$ with the same $s$ are connected by  straight deformed rulings. Assume the conditions and definitions of Theorem \ref{theorem2} on $\bfy_i(s)$. 
Let $\bfb_i(s)$ be the binormal vector of the $i^{\rm th}$ crease at $\bfy_i(s)$, and let $\bfP_i(s)=\bfI-2\bfb_i(s)\otimes\bfb_i(s)$. Then a necessary and sufficient condition for getting a compatible curved origami is
\beq
\bfP_1(s)\bfP_2(s)\dots\bfP_{n-1}(s)\bfP_n(s)=\bfI, \quad n\; {\rm is \;even}.\label{theorem5.1}
\eeq
\end{theorem}
See proof in \ref{A6}. As $n=4$, we obtain the generalized Miura origami. Let the preimage of $\bfy_i(s)$ be $\bfx_i(s),\ i=1,2,3,4,\ 0<s<s_1$.
Then the points $\bfx_1(s),\bfx_2(s),\bfx_3(s),\bfx_4(s)$ with same $s$ will lie in the same loop. Figure \ref{fig:four_fold}(a) illustrates the reference configuration of four-fold curved origami, where the gray lines represent the reference rulings. According to Corollary \ref{theorem3}, if adjacent normal vectors on both sides of each deformed crease satisfy (\ref{n_condtion}), the origami at each crease is developable and the compatibility condition (\ref{theorem5.1}) is automatically satisfied, see also \ref{A6}. Based on the assumptions in Corollary \ref{theorem3}, one algorithm to construct curved Miura origami with four creases is shown in Algorithm 6. 
\begin{breakablealgorithm}
\caption{Curved origami with multiple creases} 
\hspace*{0.02in} {\bf Input:} 
normal vectors $\bfn_1(s),\bfn_2(s),\bfn_3(s),\bfn_4(s)$.\\
\hspace*{0.02in} {\bf Output:} curved origami with multiple creases.\\
\hspace*{0.02in} {Steps:}
\begin{algorithmic}[1]
\State{Find the deformed creases:
 $\bfy_1(s),\bfy_2(s),\bfy_3(s),\bfy_4(s)$:
\beqs
\bfy_1(s)=\int_{0}^sc_1\frac{\bfn_1\times\bfn_2}{|\bfn_1\times\bfn_2|}dr,&&\bfy_2(s)=\int_{0}^sc_2\frac{\bfn_2\times\bfn_3}{|\bfn_2\times\bfn_3|}dr,\\
\bfy_3(s)=\int_{0}^sc_3\frac{\bfn_3\times\bfn_4}{|\bfn_3\times\bfn_4|}dr,&&\bfy_4(s)=\int_{0}^sc_4\frac{\bfn_4\times\bfn_1}{|\bfn_4\times\bfn_1|}dr.
\eeqs
where $\bfy_i(0)=0$ and $c_i\in C^1(0,s_1),\ i=1,2,3,4$.}
\State{ Construct the curved origami:
\beqs
\calS=\left\{\begin{array}{ll}
(1-v)\bfy_1(s)+v\bfy_2(s)\\
(1-v)\bfy_2(s)+v\bfy_3(s)\\
(1-v)\bfy_3(s)+v\bfy_4(s)\\
(1-v)\bfy_4(s)+v\bfy_1(s)
\end{array}:0<v<1,\,0<s<s_2\right\}
\eeqs
}
\end{algorithmic}
\end{breakablealgorithm}
Figure \ref{fig:four_fold}(b) shows some examples of curved Miura origami in the folded state.  
\begin{figure}[ht]
\centering
\includegraphics[scale=1]{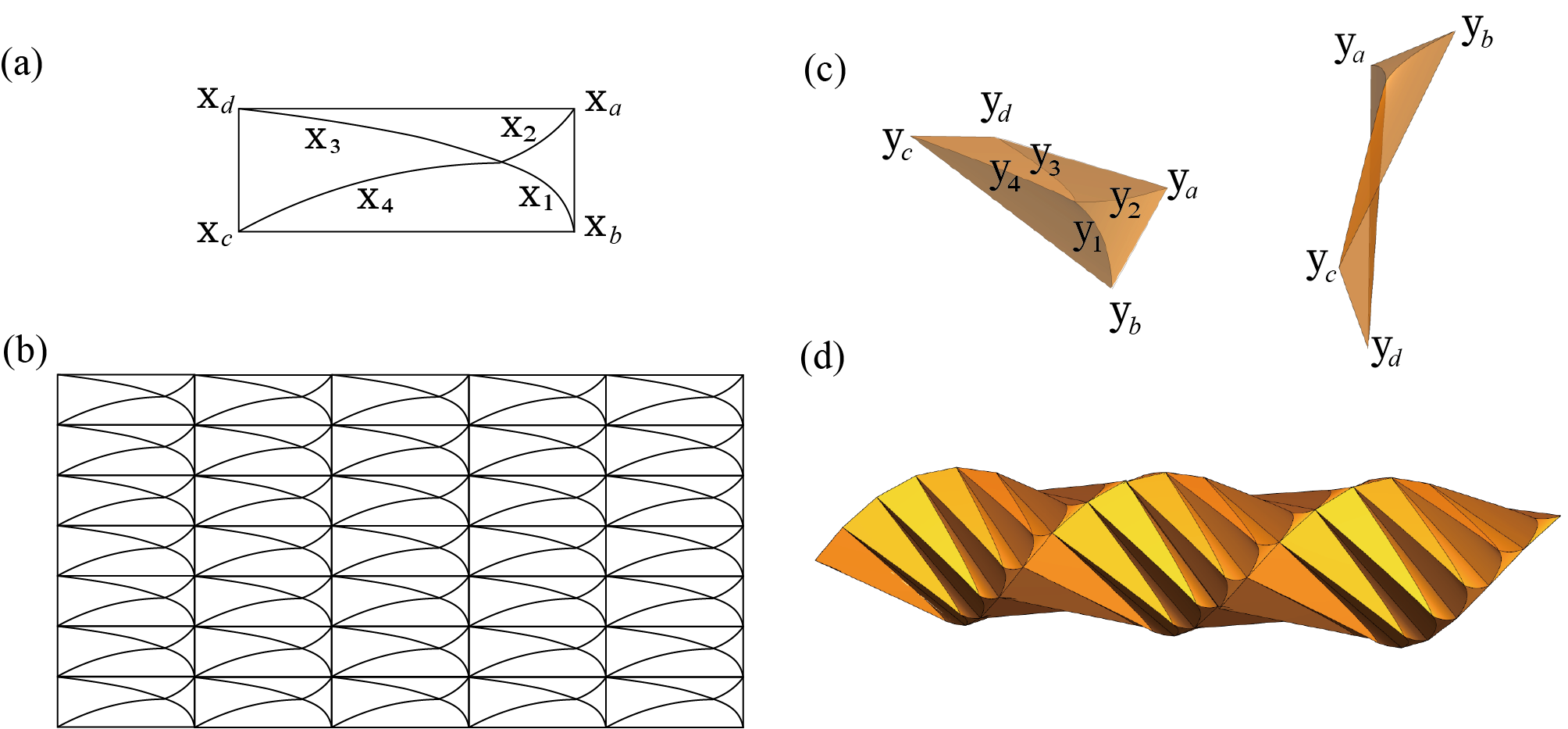}
\caption{Helical origami with four curved creases in the unit cell. (a) Reference unit cell. (b) Reference configuration. (c) Deformed unit cell in different viewpoints. (d) Deformed configuration.}
\label{fig:4fold_helical}
\end{figure}

The group orbit procedure can also be applied to the generalized Miura pattern to design complex curved origami. Figure \ref{fig:4fold_helical} shows an example in which the reference unit cell is a rectangle. We apply the translation group to the reference unit cell (Figure \ref{fig:4fold_helical}(a)) and the helical group to the deformed unit cell (Figure \ref{fig:4fold_helical}(b)), as the examples shown in Section \ref{helical_group}. In Figure \ref{fig:4fold_helical}, $\bfy_{1,2,3,4}$ are the images of the reference creases $\bfx_{1,2,3,4}$. Since isometric mappings always shorten distances, the mountain-valley assignments of the creases in this generalized Miura pattern will be 3 mountains$-$1 valley or 3 valleys$-$1 mountain, which are the same as that of the traditional Miura pattern, one can get the idea by
linearization. In this example, creases $\bfy_1,\bfy_2,\bfy_4$ are mountains and crease $\bfy_3$ is a valley. The details of the group orbit procedure will not be repeated here.

\section{Energy of curved tiles}
\subsection{Kirchhoff's nonlinear plate theory}
\label{Kirchhoff's plate theory}
The energy of an isometrically deformed surface can be accurately found by Kirchhoff's nonlinear plate theory. For an isometric mapping $\hat\bfy:\Omega\subset\R^2\to\calS\subset\R^3$, the energy is \cite{friesecke2002theorem}
\beq
\calE_{_{\calS}}=\frac{h^3}{24}\int_{\Omega}2\mu|\Rmnum2|^2+\frac{\lambda\mu}{\mu+\lambda/2}(\tr\Rmnum2)^2d\bfx,\label{energy_james}
\eeq
where $h$ is the thickness of $\Omega$, $\Rmnum2=\bfn\cdot\nabla\bfF=-\bfF^{\rm T}\nabla\bfn$ is the second fundamental form, $\bfn$ is the normal vector of the surface, $\lambda,\mu$ are the Lam\'e moduli. Substituting (\ref{Eq:gradients2}), the second fundamental form becomes
\beq
\Rmnum2=\Lambda\bfe^{\perp}\otimes\bfe^{\perp}.
\eeq
Thus, (\ref{energy_james}) can be further simplified to
\beq
\calE_{_{\calS}}=\frac{E h^3}{24}\int_{\Omega}\Lambda^2d\bfx,
\eeq
where $E=2\mu+\frac{\lambda\mu}{\mu+\lambda/2}$ is the plate modulus. The reference configuration $\Omega$ can be parameterized as $\bfx(s,v) = v \bfe(s) + \bfx_0(s),s_1<s<s_2,v_1(s)< v< v_2(s)$ by (\ref{ref_rule}). So we can get
\beq
\calE_{_{\calS}}=\frac{E h^3}{24}\int_{s_1}^{s_2}\int_{v_1}^{v_2}\Lambda^2Jdvds,\label{energy_lamda}
\eeq
where the Jacobian determinant $J$ is
\beq
J=\frac{\partial\bfx(x_1,x_2)}{\partial\bfx(s,v)}=\det \left( \begin{array}{cc} \bfx'\cdot\hat\bfe_1 & \bfe\cdot\hat\bfe_1 \\
\bfx'\cdot\hat\bfe_2 & \bfe\cdot\hat\bfe_2 \end{array}  \right)=\bfx'\cdot\bfe^\perp,
\eeq
and $(\hat\bfe_1,\hat\bfe_2)$ is a fixed orthonormal basis with associated coordinates $(x_1,x_2)$. 
Since $\bfn'=\nabla\bfn\cdot\bfx'$ and $\Rmnum2=-\bfF^{\rm T}\nabla\bfn$, we get
\beq
\Lambda^2=\frac{|\bfn'|^2}{(\bfx'\cdot\bfe^\perp)^2}.
\eeq
Thus, the energy $\calE_{_\calS}$ is
\beq
\calE_{_\calS}=\frac{E h^3}{24}\int_{s_1}^{s_2}\int_{v_1}^{v_2}\frac{|\bfn'|^2}{\bfx'\cdot\bfe^\perp}dvds.
\label{eq:isometric_energy}
\eeq
Then substituting $\bfx'=\bfx_0'+v\bfe'$, we get
\beq
\calE_{_\calS}=\left\{\begin{aligned}
&\frac{E h^3}{24}\int_{s_1}^{s_2}\frac{|\bfn'|^2}{\bfe'\cdot\bfe^\perp}\log(\frac{\bfx_0'\cdot\bfe^\perp+v_2\bfe'\cdot\bfe^\perp}{\bfx_0'\cdot\bfe^\perp+v_1\bfe'\cdot\bfe^\perp})ds,& \bfe'\neq0, \\
&\frac{E h^3}{24}\int_{s_1}^{s_2}\frac{|\bfn'|^2}{\bfx_0'\cdot\bfe^\perp}(v_2-v_1)ds,& \bfe'=0. \end{aligned}\right.\label{total_energy0}
\eeq
For a curved origami structure, $\bfx_0(s)$ is the reference crease and for each tile, we have $0<v<\tilde v(s)$, where $\tilde v(s)$ is the length of the reference ruling at position $s$. Then the energy of each tile can be given by
\beq
\calE_{_\calS}=\left\{\begin{aligned}
&\frac{E h^3}{24}\int_{s_1}^{s_2}\frac{|\bfn'|^2}{\bfe'\cdot\bfe^\perp}\log(\frac{\bfe'\cdot\bfe^\perp}{\bfx_0'\cdot\bfe^\perp}\tilde v+1)ds,& \bfe'\neq0, \\
&\frac{E h^3}{24}\int_{s_1}^{s_2}\frac{|\bfn'|^2\tilde v}{\bfx_0'\cdot\bfe^\perp}ds,& \bfe'=0. \end{aligned}\right.
\label{total_energy}
\eeq
We also discuss the energy of the case where $\bfx_0'\cdot\bfe^\perp=0$, i.e., the surface is a tangent surface, see details in \ref{tangent_surface}.

In summary, for curved tile origami, the
elastic energy according to Kirchhoff theory
written using the parameterization given in this paper can be expressed as a simple, explicit
1-dimensional integral.

\subsection{Discussion of the folding motion and the energy landscape}
In this section, we discuss one strategy to get folding motions of curved origami. Different from rigid-foldable origami whose creases and tiles undergo rigid-body motions \cite{feng2020designs}, curved origami with flexible creases and tiles exhibits infinite degrees of freedom of deformation during folding. One way to get a folding motion of curved origami is to prescribe the deformation process of its creases. Based on Theorem \ref{theorem2}, if the reference crease and the deformed crease are given, the deformed configuration of curved origami will be determined. By designing the deformation process of the crease from $\bfx_0$ to $\bfy_0$ and ensuring the crease at each intermediate state satisfy the assumptions in the theorem, one will find the origami structure at each step and therefore get a folding motion. 

\begin{figure}[ht]
\centering
\includegraphics[scale=1]{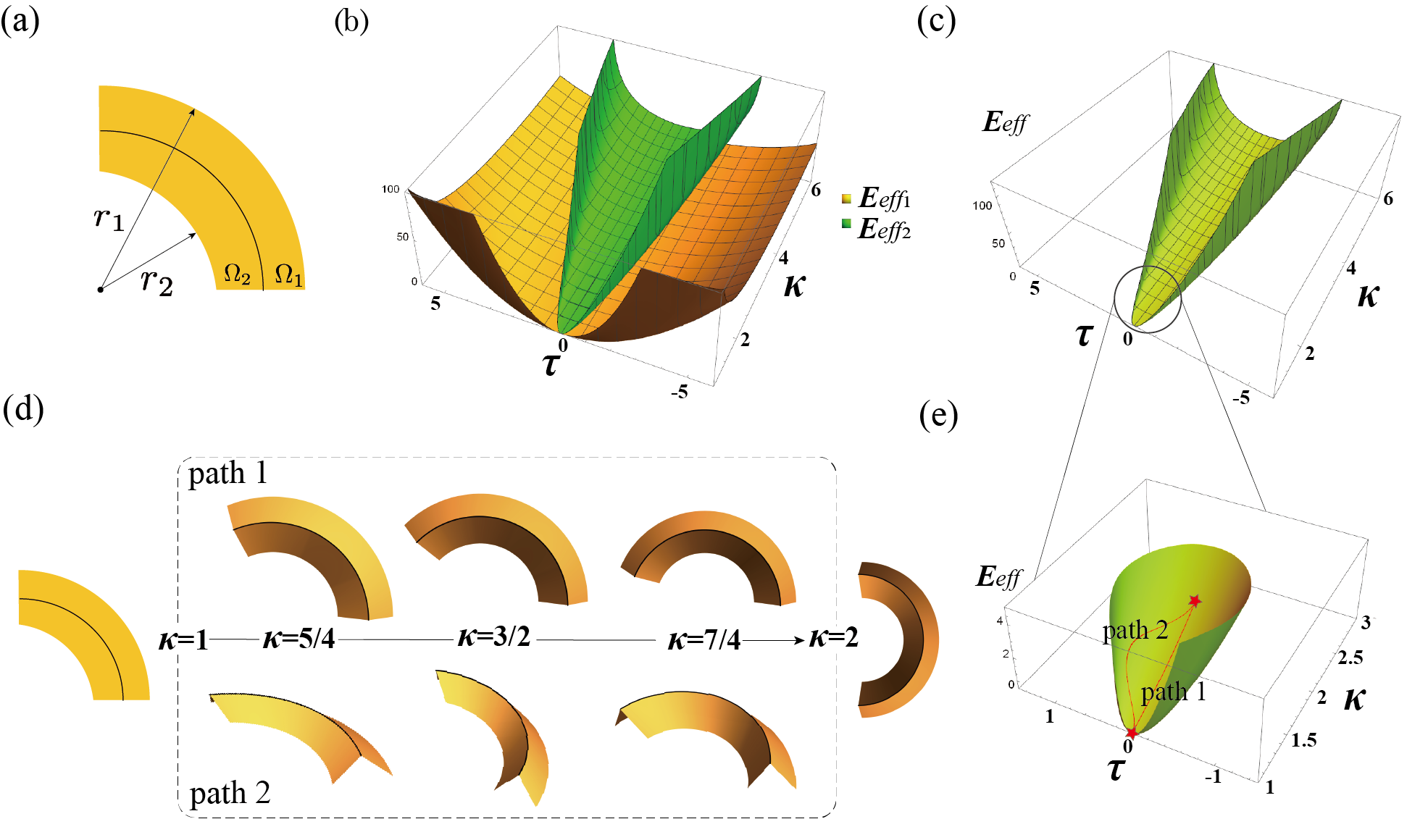}
\caption{An example to show the folding process of curved tile origami. (a) The reference domain. (b) The energy landscapes of the two surfaces. (c) The total energy stored in the curved origami. (d) Two folding motions. (e) The corresponding energy paths of the two folding motions of (d).}
\label{fold_paths}
\end{figure}

Here is an example to show the strategy. As shown in Figure \ref{fold_paths}(a), the reference crease $\bfx_0(s)$, the deformed crease $\bfy_0(s)$ and the two reference regions $\Omega_1,\Omega_2$ on the opposite sides of $\bfx_0(s)$ are given by
\beqs
\bfx_0(s)&=&\cos s \bfe_1+\sin s \bfe_2,\quad s\in(0,\pi/2),\\
\bfy_0(s)&=&(\cos2 s\bfe_1+\sin2 s\bfe_2)/2,\quad s\in(0,\pi/2),\\
\Omega_1&=&\{r\cos s\bfe_1+r\sin s\bfe_2:s\in(0,\pi/2),r\in(1,3/2)\},\\
\Omega_2&=&\{r\cos s\bfe_1+r\sin s\bfe_2:s\in(0,\pi/2),r\in(1/2,1)\}.
\eeqs
Since a curve can be uniquely defined by its curvature and torsion \cite{kreyszig2019introduction}, we use the curvature $\kappa$ and torsion $\tau$ to parameterize the crease during folding. Let $\kappa_t(s)$ and $\tau_t(s)$ denote the curvature and torsion of the crease at arc length $s\in(s_1,s_2)$ at time $t\in[0,1]$ respectively, where $t=0$ represents the reference state and $t=1$ represents the final state.  Then in this example, \beq
\kappa_0(s)=|\bfx_0''|=1,\quad \kappa_1(s)=|\bfy_0''|=2,\quad\tau_0(s)=\tau_1(s)=0.
\eeq
To simplify the problem, we assume $\kappa$ and $\tau$ are independent of $s$ during folding. The crease at time $t$ can be expressed as
\beq
\bfy_0^{t}(s)=\frac{\kappa_t}{\omega_t^2}\cos(\omega_t s)\hat\bfe_1+\frac{\kappa_t}{\omega_t^2}\sin(\omega_t s)\hat\bfe_2+\frac{\tau_t s}{\omega_t}\hat\bfe_3, \ s_1<s<s_2
\eeq
where $\omega_t=\sqrt{\kappa_t^2+\tau_t^2}$ and $\kappa_t\geq1$, $(\hat\bfe_1,\hat\bfe_2,\hat\bfe_3)$ is a fixed orthornormal basis in $\R^3$. By continuously changing $\kappa_t$ and $\tau_t$ from $(\kappa_0,\tau_0)=(1,0)$ to $(\kappa_1,\tau_1)=(2,0)$, one will get deformation process of the crease and the associated folding motion the whole origami structure. Two folding motions are shown in Figure \ref{fold_paths}(d), the corresponding curvature and torsion in paths 1 and 2 are
\beq
\left\{\begin{array}{ll}
\kappa_t=t+1,\\
\tau_t=0,
\end{array}\right.{\rm and}\quad
\left\{\begin{array}{ll}
\kappa_t=t+1,\\
\tau_t=7t^2(t-1)^2.
\end{array}\right.
\eeq

The energy of the two tiles $\calE_{_{\calS_1}}$ and $\calE_{_{\calS_2}}$ in terms of curvature and torsion can be found by using Kirchhoff's nonlinear plate theory in (\ref{total_energy}), which is 
\beq
\calE=\calE_{_{\calS_1}}+\calE_{_{\calS_2}}
=\frac{\pi Eh^3}{48}E_{eff},
\eeq
where
\beq
E_{eff}=E_{eff_1}+E_{eff_2}=\frac{1}{2}(\kappa^2+\tau^2-1)\log\frac{r_1^2(\kappa^2+\tau^2-1)-\tau^2}{r_2^2(\kappa^2+\tau^2-1)-\tau^2}\label{Eeff}
\eeq
and $E_{eff_1},E_{eff_2}$ are the effective energy of $\calS_1,\calS_2$. Here we drop the subscript $t$ in (\ref{Eeff}) and in Figure \ref{fold_paths}. The energy landscapes of $E_{eff_1},E_{eff_2}$ and $E_{eff}$ are shown in Figure \ref{fold_paths}(b-c). The two red stars in Figure \ref{fold_paths}(e) correspond to the reference state and the final deformed state, and the two folding motions are highlighted by red lines. Without considering the intersection between $\calS_1$ and $\calS_2$, each point in the energy surface in Figure \ref{fold_paths}(c) will be a physical folding state, and therefore any $C^0$ curve in the energy surface of $E_{eff}$ will correspond to a folding motion. 

\begin{figure}[ht]
\centering
\includegraphics[scale=1]{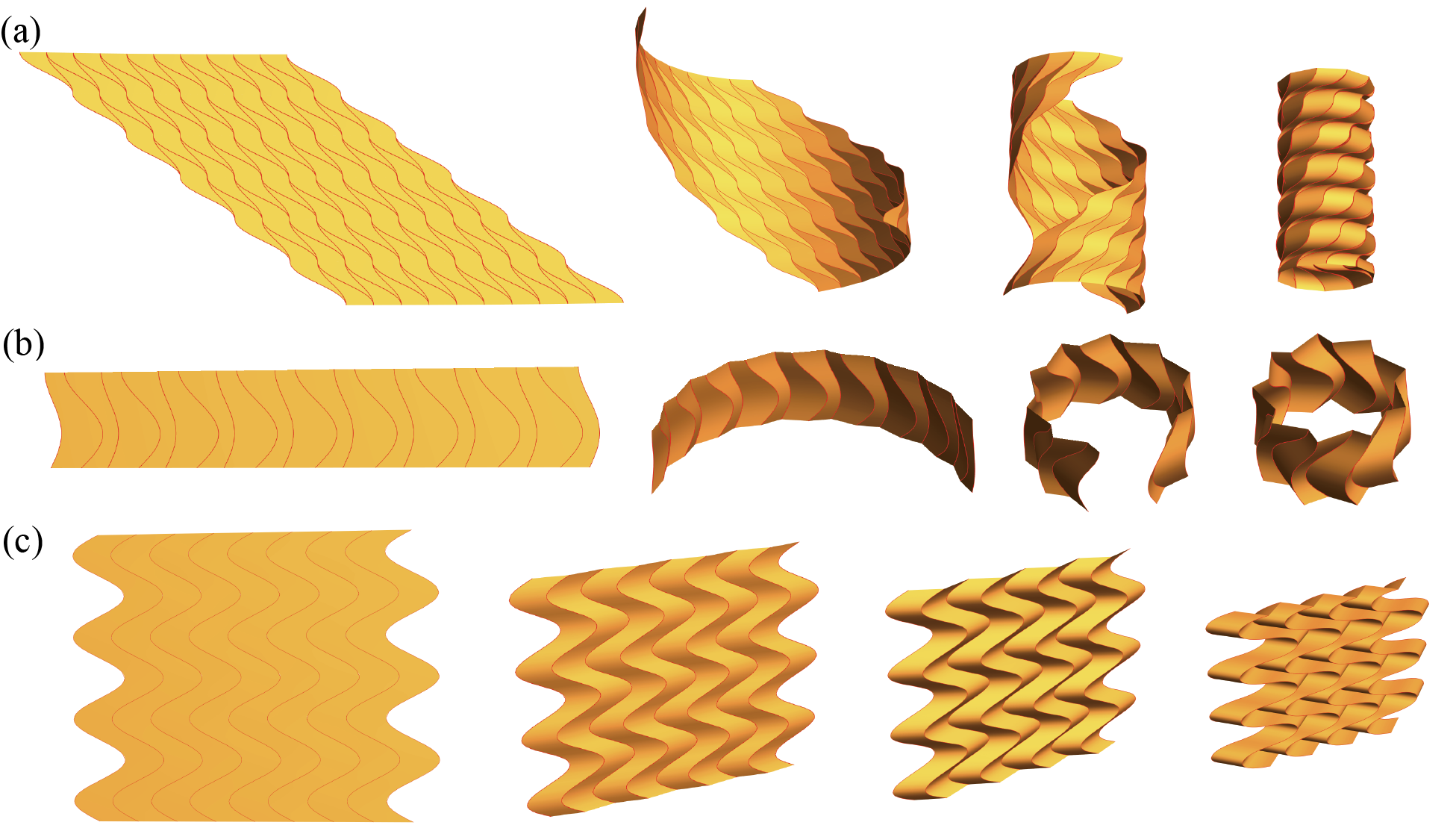}
\caption{Snapshots of curved origami structures during folding in which (a) helical groups, (b) circle groups, and (c) translation groups are applied at each stage in the folding processes. }
\label{folding_motion}
\end{figure}

From above example, the folding motions of curved origami with one crease can be obtained by designing the deformation process of its crease. For origami structures shown in Section \ref{group_orbit}, by specifying a suitable deformation process of the crease in the unit cell, group orbit procedure is potentially applied to the unit cell at each intermediate state, and then we will get a folding motion of the whole origami structure. Figure \ref{folding_motion} shows three examples of folding motions in which helical groups, circle groups and translation groups are applied respectively. See animations of the three examples in Movie 1,2 and 3.

In some applications, one would like a flat 
sheet to fold up spontaneously to one of 
the deformed structures shown in this paper.
This is possible using our results and, for example, a thin sheet of the well-known Ni$_{50.6}$Ti$_{49.4}$ shape memory alloy
having martensite as the stable room temperature phase. 
For this purpose one would begin with a flat stress-free sheet of NiTi at room temperature and deform the sheet
isometrically into the desired shape, e.g., one of the deformed shapes shown in this paper.  Then, by using a suitable fixture, the sample would be held in that shape.
Next, one would apply the standard shape-setting heat treatment for NiTi to this 
deformed sheet. After cooling to room temperature, removing the fixture and flattening the specimen, heating would cause the specimen to spontaneously deformed to the folded state, assuming the existence of an energy-decreasing path.  A similar but
more sophisticated method involving also a preliminary
selective etching of creases has been
demonstrated in \cite{velvaluri2021origami}. With suitable
measured material properties, the methods of this section are applicable to these cases.

\section{Buckling patterns of thin-walled cylindrical and conical shells} \label{buckling_pattern}

In this section we suggest one way to characterize the buckled geometry observed
in cylindrical shells and conical shells. In the early work on cylindrical shell buckling \cite{horton1965imperfections,seffen2014surface}, workers fitted a thin-walled cylinder onto a mandrel core with a prescribed annular gap, and then compressed the cylinder axially to get a surface texture of diamond-shaped buckles. The Yoshimura origami pattern with straight creases is often used to characterize this buckled geometry \cite{hunt2005twist}. From the experiments shown in \cite{horton1965imperfections,seffen2014surface,lundquist1934strength}, we notice that the buckle edges in most patterns are curves rather than straight segments, and the overall layout of buckles is periodic in the axial direction and has rotational symmetry. This structure resembles quite closely our curved origami shown in the bottom left of Figure \ref{rotation}. So, curved origami is potentially useful to describe 
accurately the buckling patterns. The strategy of constructing such origami structures is given in Algorithm \ref{trans_circle}. Since the accurate profiles of the buckle edges are not clear, the creases we prescribed to reconstruct buckling patterns are based on our observation. One can choose other curves to better fit the buckle edges according to their experiments.
\begin{figure}[ht]
\centering
\includegraphics[scale=1]{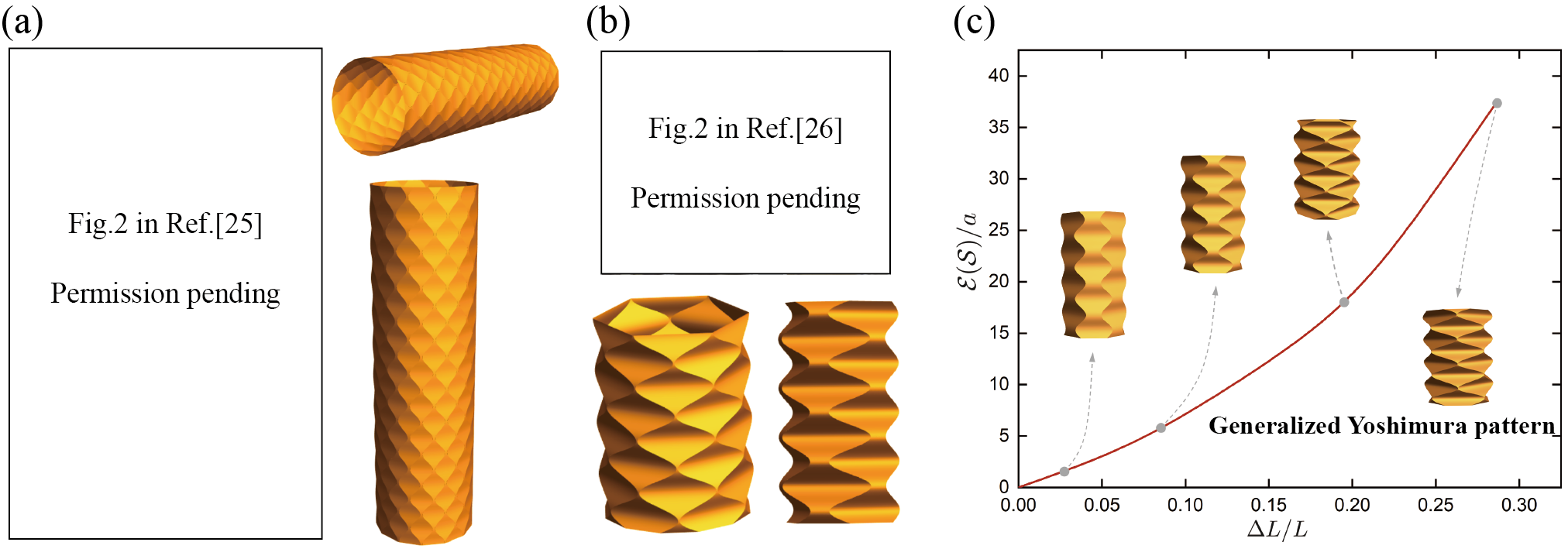}
\caption{(a) Left: bucking pattern in \cite{horton1965imperfections}; Right: curved origami structure resembling the buckling pattern on the left. (b) Top: bucking patterns in \cite{seffen2014surface}, where the bumps in the dimples are caused by the mandrel core inside the cylinder; Bottom: curved origami structures resembling the buckling patterns on the top. (c) The stored energy of different buckling patterns of the same thin-walled cylindrical shell, where $a\in\R$ is a constant, $L$ represents the original height of the cylindrical shell and $\Delta L$ represents the height of shortening after buckling. }
\label{buckling}
\end{figure}

In Figure \ref{buckling}(a) and (b), we rebuild some buckling patterns in \cite{horton1965imperfections,seffen2014surface} with origami structures. Here periodic trigonometric functions are used to characterize the profiles of buckles, which later become the creases of the origami structures. 
By varying the period and amplitude of the trigonometric function and the number of the creases, one can construct different buckling patterns.

Assuming elasticity theory applies, the energy stored in the buckled surfaces can be found by using Kirchhoff's plate theory.  According to (\ref{energy_lamda}), the energy in the buckled cylinder can be found by
\beq
    \calE(\calS)=\frac{\pi r E h^3}{12}\int_{0}^{H}\Lambda(x)^2dx
\eeq
where $r$ is the radius of the original cylinder, $H$ is the height of the buckled cylinder, and $\Lambda(x)$ is the curvature of the surface curve exactly halfway between two adjacent creases.  Figure \ref{buckling}(c) shows the stored energy of different buckling patterns of the same cylindrical shell versus $\Delta L/L$, where $L$ represents original height of the cylindrical shell, $\Delta L=L-H$, and $a=\frac{\pi r E h^3}{12}$. In this plot, each value of $\Delta L/L$ will correspond to a unique buckling pattern since we fix the number of creases and the number of periods of each crease. One can see that the greater the buckling, the more energy is stored. It's worth mentioning that as the degree of the buckling increases, the two adjacent creases will touch
at the peaks, and we will get a generalized Yoshimura pattern (see traditional Yoshimura pattern in \cite{hunt2005twist}, where the creases are straight). Note that these curved origami structures are rigid, which is consistent with the observation in experiments. They have different reference crease patterns, not intermediate states of a continuous folding of one crease pattern.

\begin{figure}[H]
\centering
\includegraphics[scale=1]{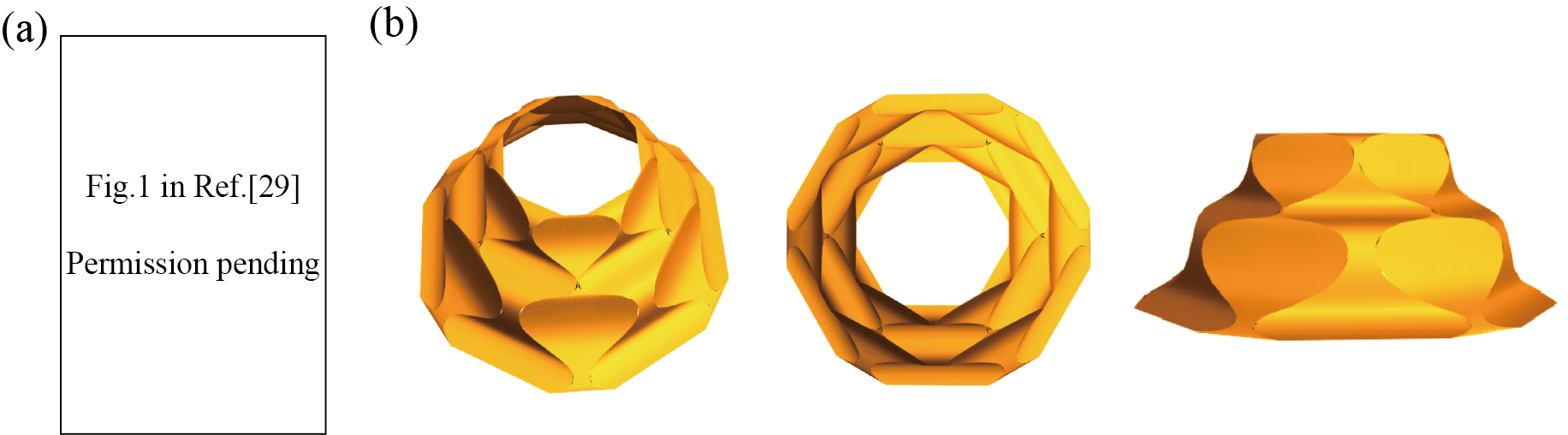}
\caption{(a) Bucking pattern of the thin-walled conical shell in \cite{weingarten1965elastic}. (b) Different viewpoints of curved origami structures resembling the buckling pattern of (a).}
\label{buckling_cone}
\end{figure}

Thin-walled conical shells under uniaxial compression also exhibit diamond-shaped buckles similar to the cylindrical shell case, but the size of the buckles increases gradually from the small radius end to the large radius end \cite{weingarten1965elastic}. Curved origami also resembles the buckling patterns in this case, and Algorithm \ref{circle_groups} is used for the construction since both reference configuration and deformed configuration have rotational symmetry. Figure \ref{buckling_cone} shows an example to resemble the buckling pattern in \cite{weingarten1965elastic}. 

\vspace{2mm}
\noindent {\bf Acknowledgments} The authors acknowledge funding from a Vannevar Bush Faculty  Fellowship to RDJ (ONR Grant No. N00014-19-1-2623).  The work also benefited from the support of
AFOSR (Grant No. FA9550-23-1-0093).

\appendix
\section{Appendix}
\subsection{Proof of Theorem \ref{multiple_creases}}
\label{A6}
\begin{proof}
We first prove the necessity. Let the reference creases be $\bfx_i(s),s_1<s<s_2 ,1=1,2,\dots,n,n>2$, which are arranged counterclockwise. Let image of $\bfx_i(s)$ be $\bfy_i(s)$. Define ruling $\bft_i$, normal $\bfn_i$ and $\bft_i^\perp$ of $i^{\rm th}$ surface bounded by $\bfy_i(s)$ and $\bfy_{i+1}(s)$ by
\beq\bft_i=\frac{\bfy_{i+1}-\bfy_i}{|\bfy_{i+1}-\bfy_i|},\;\bfn_i=\frac{\bfy_i'\times\bft_i}{|\bfy_i'\times\bft_i|},\;\bft_i^\perp=\bfn_i\times\bft_i,\; i=1,\dots,n,
\eeq
where $\bfy_{n+1}=\bfy_1$. $\bft^\perp_i$ defined above ensures that 
\beq
\bft_{i-1}^\perp\cdot\bfy_i'<0\quad\bft_{i}^\perp\cdot\bfy_{i}'<0.\eeq
At crease $\bfy_{i}$, we have $\bfn_{i-1}\cdot\bfy_{i}''=-\bfn_{i}\cdot\bfy_{i}'',i=2,\dots,n$ by (\ref{developability}). Since $(\bfn_{i-1}\cdot\bfy_i')'=(\bfn_{i}\cdot\bfy_i')'=0$, we get $\bfn_{i-1}'\cdot\bfy_i'=-\bfn_i'\cdot\bfy_{i}'$, that is,
\beq
(\bfn_{i-1}'\cdot\bft_{i-1}^\perp)(\bft_{i-1}^\perp\cdot\bfy_i')=-(\bfn_{i}'\cdot\bft_{i}^\perp)(\bft_{i}^\perp\cdot\bfy_{i}'),\quad i=2,\dots,n.\label{n_i_i-1}
\eeq
Thus, if $\bfn_{i}'\neq0$, sign$(\bfn_{i-1}'\cdot\bft_{i-1}^\perp)=-$sign$(\bfn_{i}'\cdot\bft_{i}^\perp)$. This means 
\beq
{\rm sign}(\bfn_1'\cdot\bft_1^\perp)=(-1)^{n-1}{\rm sign}(\bfn_{n}'\cdot\bft_{n}^\perp)=-{\rm sign}(\bfn_{n}'\cdot\bft_{n}^\perp).
\eeq
Thus, $n$ is even.

Let $\bfF_i(s)$ be the deformation gradient of the ruling between $\bfy_i(s)$ and $\bfy_{i+1}(s)$. According to Corollary \ref{theorem3} and Corollary \ref{normal_grad},  the deformation gradients will satisfy
\beq
\bfF_1=\bfP_1\bfF_2,\quad\bfF_2=\bfP_2\bfF_3,\quad\dots,\quad\bfF_{n-1}=\bfP_{n-1}\bfF_{n},\quad\bfF_{n}=\bfP_{n}\bfF_{1}.
\eeq
where $\bfP_i=\bfI-2\bfb_i\otimes\bfb_i$. Thus, we get 
\beq
\bfF_1=\bfP_1\bfP_2\dots\bfP_n\bfF_1.
\eeq
Since $\bfP_1\bfP_2\dots\bfP_n\in$ SO(3), we have
\beq
\bfP_1\bfP_2\dots\bfP_n=\bfI.\label{odd_even_reflections}
\eeq
Therefore, $\bfP_1\bfP_1\dots\bfP_n=\bfI$ and $n$ is even.

Then we show sufficiency. The deformation gradients of the $n$ surfaces are given by
\beq
\nabla\hat\bfy=\left\{\begin{array}{lc}
    \bfF_1, &  {\rm if}\ \bfx= (1-\lambda)\bfx_1+\lambda\bfx_2\\
    \bfP_2\bfF_1, & {\rm if}\ \bfx=  (1-\lambda)\bfx_2+\lambda\bfx_3\\
     \bfP_3\bfP_2\bfF_1, & {\rm if}\ \bfx= (1-\lambda)\bfx_3+\lambda\bfx_4\\
     \dots &\dots \\
      \bfP_{n}\dots\bfP_3\bfP_2\bfF_1, & {\rm if}\ \bfx= (1-\lambda)\bfx_n+\lambda\bfx_1\\
\end{array}\right.
\eeq
where $0<\lambda<1$. At crease $\bfx_i,i=2,\dots,n$, we have \beq
\bfy_i=\bfP_{i-1}\dots\bfP_3\bfP_2\bfF_1\bfx_i'=\bfP_i(\bfP_{i-1}\dots\bfP_3\bfP_2\bfF_1\bfx_i').
\eeq
So the compatibility condition at $i^{\rm th}$ crease is automatically satisfied,  $i=2,\dots,n$. The remaining compatibility condition is at $1$st crease. As $\bfP_n\dots\bfP_3\bfP_2\bfP_1=\bfI$ and $\bfy_1'=\bfF_1\bfx_1=\bfP_1\bfF_1\bfx_1$, we have
\beq
\bfF_1\bfx_1'=\bfP_{n}\dots\bfP_3\bfP_2\bfP_1\bfF_1\bfx_1'=\bfP_{n}\dots\bfP_3\bfP_2\bfF_1\bfx_1',
\eeq
So the $n$th surface is compatible with the $1$st surface at $1$st crease. Thus, the curved origami structure is compatible.
\end{proof}
\subsection{Energy stored in a tangent surface}
\label{tangent_surface}
If $\bfx_0'\cdot\bfe^\perp=\bfy_0'\cdot\bft^\perp=0$,  the developable surface is a tangent surface. Then the rulings and normals are found by 
\beqs
\bfe&=&\bfx_0',\quad\bft=\bfy_0',\\
\bfe'&=&\bfx_0'',\quad\bfe^\perp=\bfp_0,\\
\bft'&=&\bfy_0'',\quad\bft^\perp=\bfp,\\
\bfn&=&\bft\times\bft^\perp=\bfb.
\eeqs
So $\bfx_0''\cdot\bfp_0=\bfy_0''\cdot\bfp=\bfe'\cdot\bfe^\perp=\bft'\cdot\bft^\perp$. From (\ref{total_energy0}) we have
\beqs
\calE_{_\calS}&=&\frac{E h^3}{24}\int_{s_1}^{s_2}\frac{|\bfn'|^2}{\bfe'\cdot\bfe^\perp}\log(\frac{v_2}{v_1})ds,\nonumber\\
&=&\frac{E h^3}{24}\int_{s_1}^{s_2}\frac{|\bfb'|^2}{\bfy_0''\cdot\bfp}\log(\frac{v_2}{v_1})ds,\nonumber\\
&=&\frac{E h^3}{24}\int_{s_1}^{s_2}\frac{\tau^2}{\kappa}\log(\frac{v_2}{v_1})ds.
\eeqs

\bibliography{reference}
\end{document}